\documentclass[a4paper,12pt, epsfig]{article}
\pdfoutput=1
\usepackage{epsfig}
\usepackage{epstopdf}
\usepackage{graphicx}
\usepackage{ifthen}
\usepackage{feynmp}
\usepackage{amsmath}
\usepackage[psamsfonts]{amssymb}
\usepackage{euscript}

\usepackage{latexsym}
\usepackage[arrow,matrix,curve]{xy}

\newskip\humongous \humongous=0pt plus 1000pt minus 1000pt

\newif\ifdtup

\def\,{\hspace{-.1cm}}
\def\hsp{,\hspace{.7cm}}

\def\fc#1#2 {\frac{n}{q}#1\frac{n}{q}#2}

\def\vpp{V^{\prime\prime}}
\def\vppp{V^{\prime\prime\prime}}

\newcommand{\vac}{\ensuremath{|0\rangle}}

\renewcommand{\cos}{\textrm{cos}}
\renewcommand{\sin}{\textrm{sin}}

\renewcommand{\sinh}{\textrm{sinh}}
\renewcommand{\cosh}{\textrm{cosh}}
\renewcommand{\tanh}{\textrm{tanh}}
\newcommand{\sech}{\textrm{sech}}
\newcommand{\csch}{\textrm{csch}}

\def\exp#1{\hbox{\rm exp}\left(#1\right)}

\renewcommand{\theequation}{\arabic{section}.\arabic{equation}}
\renewcommand{\(}{\begin{equation}}
\renewcommand{\)}{end{equation} \vspace{-.05in}\linebreak}

\newcounter{saveeqn}
\newcounter{savealpheqn}

\newcommand{\alpheqn}{\setcounter{saveeqn}{\value{equation}}%
  \stepcounter{saveeqn}\setcounter{equation}{0}%
  \renewcommand{\theequation}{\mbox{\arabic{section}.\arabic{saveeqn}
\alph{equation}}}
  \renewcommand{\)}{\end{equation}}}
\def\part#1{\frac{\partial}{\partial{#1}}}%
\def\group#1{\refstepcounter{equation}\setcounter{saveeqn}
 {\value{equation}}%
  \label{#1}\setcounter{equation}{0}%
\renewcommand{\theequation}{\mbox{\arabic{section}.\arabic{saveeqn}
\alph{equation}}}
  \renewcommand{\)}{\end{equation}}}
\newcommand{\reseteqn}{\setcounter{equation}{\value{saveeqn}}%
  \renewcommand{\theequation}{\arabic{section}.\arabic{equation}}%
  \renewcommand{\)}{\end{equation}}}

\newcommand{\aalpheqn}{\setcounter{saveeqn}{\value{equation}}%
  \stepcounter{saveeqn}\setcounter{equation}{0}%
  \renewcommand{\theequation}{\mbox{
        \Alph{subsection}.\arabic{saveeqn}\alph{equation}}}
   \renewcommand{\)}{\end{equation}}}
\newcommand{\areseteqn}{\setcounter{equation}{\value{saveeqn}}%
  \renewcommand{\theequation}{\Alph{subsection}.\arabic{equation}}%
  \renewcommand{\)}{\end{equation}}}

\renewcommand{\thefootnote}{\alph{footnote}}
\renewcommand{\(}{\begin{equation}}
\renewcommand{\)}{\end{equation}}
\newcommand{\ba}{\begin{eqnarray}}
\newcommand{\ea}{\end{eqnarray}}

\newcommand{\vt}{V^{(3)}[gf(x)]}
\newcommand{\cbp}{\mathop{\vtop{\ialign{##\crcr
   $\hfil\displaystyle{}\hfil$\crcr\noalign{\kern-13pt\nointerlineskip}
   \BIG{)}\hskip0pt\crcr\noalign{\kern3pt}}}}}
\newcommand{\pa}{\mathop{\vtop{\ialign{##\crcr

$\hfil\displaystyle{\oplus}\hfil$\crcr\noalign{\kern+1pt\nointerlineskip
}
   \hspace{.08in}$^{\alpha=0}$\hskip6pt\crcr\noalign{\kern3pt}}}}}
\renewcommand{\hsp}{,\hspace{.3in}}
\newcommand{\p}{^\prime}
\newcommand{\pp}{^{\prime\prime}}

\newcommand{\Z}{\ensuremath{\mathbb Z}}

\catcode`\@=11
\def\vereq#1#2{\lower3pt\vbox{\baselineskip1.5pt \lineskip1.5pt
\ialign{$\m@th#1\hfill##\hfil$\crcr#2\crcr\sim\crcr}}}
\catcode`\@=12

\renewcommand{\(}{\begin{equation}}
\renewcommand{\)}{\end{equation}}

\def\pin#1{\int \frac{d#1}{2\pi}}
\def\pink#1{\int \frac{d^{#1}k}{(2\pi)^{#1}}}
\def\pinq#1{\int \frac{d^{#1}q}{(2\pi)^{#1}}}

\def\pinkp#1{\int \frac{d^{#1}k\p}{(2\pi)^{#1}}}
\def\pinpp#1{\int \frac{d^{#1}p\p}{(2\pi)^{#1}}}
\def\sq#1#2{\sqrt{\frac{\omega_{#1}}{\omega_{#2}}}}
\def\bd#1{b^\dag_{k_{#1}}}
\def\bm#1{b_{-k_{#1}}}
\def\Bd#1{B^\dag_{k_{#1}}}

\def\df{\mathcal{D}_f}
\def\B#1{B^\dag_{k_{#1}}}
\def\Bp#1{B^\dag_{k\p_{#1}}}
\def\I{\mathcal{I}}

\newcommand{\beas}{\begin{eqnarray*}}
\newcommand{\eeas}{\end{eqnarray*}}

\newcommand{\bquo}{\begin{quote}}
\newcommand{\enqu}{\end{quote}}

\renewcommand{\Z}{{\mathbb Z}}

\def\ch{{\mathcal{H}}}
\def\co{{\mathcal{O}}}

\def\ok#1{\omega_{k_{#1}}}
\def\okp#1{\omega_{k\p_{#1}}}
\def\V#1{V^{(#1)}[gf(x)]}

\newcommand{\beq}{\begin{equation}}
\newcommand{\eeq}{\end{equation}}
\newcommand{\bea}{\begin{eqnarray}}
\newcommand{\eea}{\end{eqnarray}}

\newskip\humongous \humongous=0pt plus 1000pt minus 1000pt

\newif\ifdtup

\catcode`\@=11

\@addtoreset{equation}{section}

\def\@normalsize{\@setsize\normalsize{15pt}\xiipt\@xiipt
\abovedisplayskip 14pt plus3pt minus3pt%
\belowdisplayskip \abovedisplayskip
\abovedisplayshortskip \z@ plus3pt%
\belowdisplayshortskip 7pt plus3.5pt minus0pt}

\def\small{\@setsize\small{13.6pt}\xipt\@xipt
\abovedisplayskip 13pt plus3pt minus3pt%
\belowdisplayskip \abovedisplayskip
\abovedisplayshortskip \z@ plus3pt%
\belowdisplayshortskip 7pt plus3.5pt minus0pt
\def\@listi{\parsep 4.5pt plus 2pt minus 1pt
      \itemsep \parsep
      \topsep 9pt plus 3pt minus 3pt}}

\relax

\catcode`@=12

\catcode`\@=11

\def\section{\@startsection{section}{1}{\z@}{3.5ex plus 1ex minus  .2ex}{2.3ex plus .2ex}{\large\bf}}

\def\thesection{\arabic{section}}
\def\thesubsection{\arabic{section}.\arabic{subsection}}

\def\appendix{\setcounter{section}{0}
 \def\thesection{Appendix \Alph{section}}
 \def\thesubsection{\Alph{section}.\arabic{subsection}}
 \def\theequation{\Alph{section}.\arabic{equation}}}
\renewcommand{\theequation}{\arabic{section}.\arabic{equation}}

\begin{document}
\def\thefootnote{\fnsymbol{footnote}}
\def\thetitle{Two-Loop Scalar Kinks}
\def\autone{Jarah Evslin}
\def\auttwo{Hengyuan Guo}
\def\affa{Institute of Modern Physics, NanChangLu 509, Lanzhou 730000, China}
\def\affb{University of the Chinese Academy of Sciences, YuQuanLu 19A, Beijing 100049, China}

\begin{center}
{\large {\bf \thetitle}}

\bigskip

\bigskip

{\large \noindent  \autone{${}^{1,2}$} \footnote{jarah@impcas.ac.cn} and \auttwo{${}^{1,2}$} \footnote{guohengyuan@impcas.ac.cn}}

\vskip.7cm

1) \affa\\
2) \affb\\

\end{center}

\begin{abstract}
\noindent
At one loop, quantum kinks are described by a sum of quantum harmonic oscillator Hamiltonians, and the ground state is just the product of the oscillator ground states.  Two-loop kink masses are only known in integrable and supersymmetric cases and two-loop states have never been found.  We find the two-loop kink mass and explicitly construct the two-loop kink ground state in a scalar field theory with an arbitrary nonderivative potential.   We use a coherent state operator which maps the vacuum sector to the kink sector, allowing all states to be treated with a single Hamiltonian which needs to be renormalized only once, eliminating the need for regulator matching conditions.   Our calculation is greatly simplified by a recently introduced alternative to collective coordinates, in which the kink momentum is fixed perturbatively.


\end{abstract}

%
\setcounter{footnote}{0}
\renewcommand{\thefootnote}{\arabic{footnote}}

\section{Introduction}
Quantum solitons at strong coupling are poorly understood, and yet are widely believed to be somehow responsible for confinement in Yang-Mills and QCD.  Understanding them is therefore of critical importance.  However we would like to suggest that this is premature as solitons at weak coupling are also not understood.  

Early papers on quantum solitons produced consistent results.  Beginning with the pioneering paper \cite{dhn1}, one-loop corrections to kink masses were calculated by introducing a vacuum sector and a kink sector Hamiltonian, regularizing them both, identifying the regulators and renormalizing.  In the 1970s, the regulator was a cutoff in the number of modes.  In the 1980s, authors instead calculated one-loop corrections to the masses of supersymmetric kinks, regularizing with an energy cutoff.  It was only in the following decade that Ref.~\cite{rebhan} reported that, when applied to the same kink, these two methods yielded different masses.  

The basic problem is as follows.  A theory is defined by its Hamitonian together with a regulator and renormalization scheme.  One thus expects masses to depend on these three choices.  However, once these are fixed, the theory is fixed as are all observables.  In particular, nothing may depend on an arbitrary choice of matching conditions for regulators.  At most one such inequivalent choice may be correct, but which?

Many responses to this question have since appeared in the literature.  The most common interpretation is that some regulator matching conditions give answers which are ``bad'' \cite{bad}.  Another response is that the problem is caused by linear divergences, but these may be made logarithmic by taking a derivative with respect to a mass scale and then integrating using a physical principle to fix the constant of integration \cite{nastase}.  This strategy has been successfully employed to reproduce the two-loop mass of the Sine-Gordon soliton.  However, as noted by an overlapping collection of authors in \cite{lit}, this strategy fails with some choices of boundary conditions and, more importantly, it does not shed light on which matching conditions should be allowed.  Perhaps the most interesting suggestion, proposed in Ref.~\cite{lit}, is that an ultraviolet cutoff may only be imposed if the nontrivial background itself has no effect above that cutoff.  It is an appealing physical principle, however in practice it does not entirely determine how the density of states is to be corrected.  Ultimately the authors chose this correction to reproduce the known answer, leading one to wonder just what prescription works when the answer is not already known.  Later it was proposed \cite{local} that instead the matching condition should keep the same mode density in every sector.  However the authors note that this proposal is only expected to work at one loop.

This state of affairs has motivated our program to systematically study perturbation theory about quantum solitons in a formalism with no matching conditions.  Instead, following \cite{hepp}, we introduce a nonlocal operator which maps the vacuum sector to the one soliton sector\footnote{For a computationally similar approach without the nonlocal operator, see Ref.~\cite{stuart}.}.  This allows all computations involving both sectors to be performed using the original Hamiltonian, with no need to introduce another Hamiltonian for the soliton sector.  We thus need to renormalize only once, obviating the need for regulator matching.   In Refs.~\cite{mekink,sg} this was carried out at one loop in the 1+1d $\phi^4$ and Sine-Gordon models.  At one loop these results were known as the theory is free.  The first correction to the states was reported in Ref.~\cite{me2stato}.  The present paper continues to two loops, for a general scalar kink in 1+1 dimensions.   The kink ground state and mass are found.

We begin in Sec.~\ref{initsez} with a review of our formalism.  Then we calculate the two-loop quantum ground states in two steps.  Our states are decomposed in a power series in the zero mode $\phi_0$ of the scalar field.  We refer to the constant terms in this decomposition as $\phi_0$-primaries and others as $\phi_0$-descendants.  In Sec.~\ref{transsez} we use translation invariance to fix all $\phi_0$-descendants in terms of $\phi_0$-primaries.  Next in Sec.~\ref{ssez} we use Schrodinger's equation to find the $\phi_0$-primaries.   As an application, in Sec.~\ref{msez} we present a formula for the two-loop mass correction to kinks in 1+1 dimensional scalar theories with an arbitrary potential.  In \ref{sapp} we show that the states that we have constructed indeed solve Schrodinger's equation.

\section{Review} \label{initsez}

\begin{table}
\begin{tabular}{|l|l|}
\hline
Operator&Description\\
\hline
$\phi(x),\ \pi(x)$&The real scalar field and its conjugate momentum\\
$a^\dag_p,\ a_p,\ A^\dag_p,\ A_p$&Creation and annihilation operators in plane wave basis\\
$b^\dag_k,\ b_k,\ B^\dag_k,\ B_k$&Creation and annihilation operators in normal mode basis\\
$\phi_0,\ \pi_0$&Zero mode of $\phi(x)$ and $\pi(x)$ in normal mode basis\\
$::_a,\ ::_b$&Normal ordering with respect to $a$ or $b$ operators respectively\\
\hline
\hline
Hamiltonian&Description\\
\hline
$H$&The original Hamiltonian\\
$H^\prime$&$H$ with $\phi(x)$ shifted by kink solution $f(x)$\\
$H_n$&The $\phi^n$ term in $H^\prime$\\
\hline
Symbol&Description\\
\hline
$f(x)$&The classical kink solution\\
$\df$&Operator that translates $\phi(x)$ by the classical kink solution\\
$g_B(x)$&The kink linearized translation mode\\
$g_k(x)$&Continuum normal mode or breather\\
$\gamma_i^{mn}$&Coefficient of $\phi_0^m B^{\dag n}\vac_0$ in order $i$ ground state\\
$\Gamma_i^{mn}$&Coefficient of $\phi_0^m B^{\dag n}\vac_0$ in order $i$ Schrodinger Equation\\
$V_{ijk}$&Derivative of the potential contracted with various functions\\
$Y_{ijk}$&$V_{ijk}$ divided by a sum of frequencies\\
$\I(x)$&Contraction factor from Wick's theorem\\
$p$&Momentum\\
$k_i$&The analog of momentum for normal modes\\
$\omega_k,\ \omega_p$&The frequency corresponding to $k$ or $p$\\
$\Omega_i$&Sum of frequencies $\ok{}$\\
$\tilde{g}$&Inverse Fourier transform of $g$\\
$Q_n$&$n$-loop correction to kink energy\\
\hline
State&Description\\
\hline
$|K\rangle, \ |\Omega\rangle$&Kink and vacuum sector ground states\\
$\co|\Omega\rangle$&Translation of $|K\rangle$ by $\df^{-1}$\\
$\co_n|\Omega\rangle$&Translation of $|K\rangle$ by $\df^{-1}$  at order $n$ \\
\hline

\end{tabular}
\caption{Summary of Notation}\label{notab}
\end{table}

In this section we will review the formalism for treating quantum kinks presented in Refs.~\cite{cahill76,mekink,memassa}.  Table~\ref{notab} summarizes some of our notation.


Let $\phi(x)$ and $\pi(x)$ be a Schrodinger picture real scalar field and its conjugate in 1+1 dimensions, whose dynamics are described by the Hamiltonian
\beq
H=\int dx \ch(x) \hsp
\ch(x)=\frac{1}{2}:\pi(x)\pi(x):_a+\frac{1}{2}:\partial_x\phi(x)\partial_x\phi(x):_a+\frac{M^2}{g^2}:\mathcal{V}[g\phi(x)]:_a.\label{hsq}
\eeq
Here $M$ and $g$ have dimensions of mass and action${}^{-1/2}$ respectively.  We expand in $g^2\hbar$ and set $\hbar=1$.   Also we will define the dimensionful potential
\beq
V=M^2\mathcal{V}.
\eeq
The normal ordering $::_a$ is defined below.
 
If $V$ has degenerate minima then there will be a classical kink solution
\beq
\phi(x,t)=f(x).
\eeq
In the Schrodinger picture, where we will always work, the translation operator
\beq
\df={\rm{exp}}\left(-i\int dx f(x)\pi(x)\right) \label{df}
\eeq
satisfies \cite{mekink}
\beq
:F\left[\pi(x),\phi(x)\right]:_a\df=\df:F\left[\pi(x),\phi(x)+f(x)\right]:_a \label{fident}
\eeq
where $F$ is an arbitrary functional.   This operator takes the vacuum sector to the kink sector.   In particular one may relate the ground states $|\Omega\rangle$ and $|K\rangle$ of the two respective sectors
\beq
|K\rangle=\df \co|\Omega\rangle
\eeq
using the perturbative operator $\co$.  The kink ground state $|K\rangle$ is an eigenstate of the Hamiltonian $H$ and so $\co|\Omega\rangle$ must be an eigenstate of the Hamiltonian
\bea
H\p&=&\df^{-1} H\df=Q_0+H_2+H_I\\
H_2&=&\frac{1}{2}\int dx\left[:\pi^2(x):_a+:\left(\partial_x\phi(x)\right)^2:_a+V^{\prime\prime}[gf(x)]:\phi^2(x):_a\right.]\nonumber
\eea
Here $Q_0$ is the classical mass of the solution $f(x)$ and $H_I$ contains all higher order terms in $g$.  

The free Hamiltonian $H_2$ leads to classical linear equations of motion whose constant frequency solutions are the normal modes $g(x)$ of the kink
 \beq
 \phi(x,t)=e^{-i\omega t}g(x)\hsp
V^{\prime\prime}[gf(x)]g(x)=\omega^2g(x)+g^{\prime\prime}(x). \label{eom}
 \eeq
There will be continuum solutions $g_k(x)$ labeled by an index $k$ such that\footnote{The sign of $k$ is chosen to agree with the momentum of the corresponding plane wave at $|x|>>0$.} $\omega_k=\sqrt{M^2+k^2}$, breathers and a single Goldstone mode $g_B(x)$ 
\beq
g_B(x)=\frac{1}{\sqrt{Q_0}} f^\prime(x)
\eeq
with $\omega_B=0$.  For brevity of notation, we will not distinguish between continuum solutions and breathers, and so it will be implicit that integrals over the continuous variable $k$ include a sum over the breathers.
 
We adopt the normalization conditions
\beq
\int dx g_{k_1} (x) g^*_{k_2}(x)=2\pi \delta(k_1-k_2)\hsp
\int dx |g_{B}(x)|^2=1
\eeq
and we choose the phases such that
\beq
g_k(-x)=g_k^*(x)=g_{-k}(x). \label{congc}
\eeq
We also define inverse Fourier transforms
\beq
\tilde{g}(p)=\int dx g(x) e^{ipx}
\eeq
satisfying the completeness relations
\beq
\tilde{g}_{B}(p)\tilde{g}_{B}(q)+\pin{k}\tilde{g}_k(p)\tilde{g}_{-k}(q)=2\pi\delta(p+q). \label{pcomp}
\eeq

The same quantum field and its conjugate may be expanded in terms of plane waves
\bea
\phi(x)&=&\pin{p}\frac{1}{\sqrt{2\omega_p}}\left(a^\dag_p+a_{-p}\right) e^{-ipx}\hsp
\omega_p=\sqrt{M^2+p^2}\label{pwexp}\\
 \pi(x)&=&i\pin{p}\sqrt{\frac{\omega_p}{2}}\left(a^\dag_p-a_{-p}\right) e^{-ipx}
\nonumber
\eea
or normal modes
\bea
\phi(x)&=&\phi_{B}(x) +\phi_{C}(x)\hsp
\pi(x)=\pi_{B}(x)+\pi_{C}(x) \label{nmexp}\\
\phi_B(x)&=&\phi_0 g_B(x)\hsp 
\phi_C(x)=\pin{k}\frac{1}{\sqrt{2\omega_k}}\left(b_k^\dag+b_{-k}\right) g_k(x)\nonumber\\
\pi_B(x)&=&\pi_0 g_B(x)\hsp
\pi_C(x)=i\pin{k}\sqrt{\frac{\omega_k}{2}}\left(b_k^\dag - b_{-k}\right) g_k(x).
\nonumber
\eea
We define the plane wave normal ordering $::_a$ by placing the $a^\dag$ to the left of the $a$ and normal mode normal ordering $::_b$ by placing $b^\dag$ and $\phi_0$ to the left of $b$ and $\pi_0$.

Using the canonical algebra satisfied by $\phi(x)$ and $\pi(x)$ together with the completeness of the solutions \cite{me2stato}
\beq
g_B(x)g_B(y)+\pin{k}g_k(x)g_{-k}(y)=\delta(x-y) \label{comp}
\eeq
 one finds
\bea
[a_p,a_q^\dag]&=&2\pi\delta(p-q)\hsp
[\phi_0,\pi_0]=i\hsp
[b_{k_1},b^\dag_{k_2}]=2\pi\delta(k_1-k_2).\nonumber
\eea
These allow the plane wave normal ordered $H_2$ to be rewritten in terms of a normal mode normal ordered free Hamiltonian plus a constant $Q_1$, which is the one-loop correction to the kink mass.  This can be achieved one term a time
\bea
:\pi_B^2(x):_a&=&:\pi_B^2(x):_b+g_B(x)\hat{\hat{g}}_B(x)\hsp 
\hat{\hat{g}}_B(x)=-\pin{p}e^{-ixp}\frac{\omega_p}{2} \tilde{g}_B(p)\\
:\pi_C^2(x):_a&=&:\pi_C^2(x):_b+\pin{k}g_k(x)\hat{\hat{g}}_{-k}(x)\hsp 
\hat{\hat{g}}_k(x)=\pin{p}e^{-ixp}\left(\frac{\omega_k-\omega_p}{2}\right) \tilde{g}_k(p)\nonumber\\
:\phi_B^2(x):_a&=&:\phi_B^2(x):_b+g_B(x)\hat{g}_B(x)\hsp 
{\hat{g}}_B(x)=-\pin{p}e^{-ixp}\frac{1}{2\omega_p} \tilde{g}_B(p)\nonumber\\
:\phi_C^2(x):_a&=&:\phi_C^2(x):_b+\pin{k}g_k(x)\hat{g}_{-k}(x)\hsp 
{\hat{g}}_k(x)=\pin{p}e^{-ixp}\left(\frac{1}{2\omega_k}-\frac{1}{2\omega_p}\right) \tilde{g}_k(p).\nonumber
\eea
Applying the classical equations of motion (\ref{eom}) one finds
\bea
V^{\prime\prime}[gf(x)]:\phi_B^2(x):_a&=&V^{\prime\prime}[gf(x)]:\phi_B^2(x):_b+g_B^{\prime\prime}(x)\hat{g}_{B}(x)\\
V^{\prime\prime}[gf(x)]:\phi_C^2(x):_a&=&V^{\prime\prime}[gf(x)]:\phi_C^2(x):_b+\pin{k}\left(\omega_k^2 g_k(x)+g_k^{\prime\prime}(x)\right)\hat{g}_{-k}(x).\nonumber
\eea
The $g\pp$ terms cancel $:\partial\phi(x)\partial\phi(x):_a-:\partial\phi(x)\partial\phi(x):_b$ after an integration by parts, leaving
\bea
H_2&=&Q_1+\frac{\pi_0^2}{2}+\pin{k}\omega_k b^\dag_k b_k \\
Q_1&=&-\frac{1}{4}\pin{k}\pin{p}\frac{(\omega_p-\omega_k)^2}{\omega_p}\tilde{g}^2_{k}(p)
-\frac{1}{4}\pin{p}\omega_p\tilde{g}_{B}(p)\tilde{g}_{B}(p).\nonumber
\eea
We perform a semiclassical expansion of the kink ground state\footnote{The $n$-loop ground state is the sum up to $i=2n-2$.  Note that there is no tree level term.  In a sense made precise in Ref.~\cite{me2stato}, the tree level ground state $|\Omega\rangle$ is automatically included in the one-loop $\vac_0$ by the condition (\ref{v0}).  }  in powers of $g$
\beq
\co|\Omega\rangle=\sum_{i=0}^\infty |0\rangle_{i}. \label{semi}
\eeq
The one-loop kink ground state $\vac_0$ is a product of free vacua 
\beq
\pi_0\vac_0=b_k\vac_0=0. \label{v0}
\eeq

In Ref.~\cite{wick} we found a general Wick's formula for the conversion of plane wave to normal mode normal ordering.  For powers of $\phi(x)$ it reads
\beq
:\phi^n(x):_a=\sum_{m=0}^{\lfloor\frac{n}{2}\rfloor}\frac{n!}{2^m m!(n-2m)!}\I^m(x):\phi^{n-2m}(x):_b
\eeq
where
\bea
\I(x)&=&g_B(x)\hat{g}_B(x)+\pin{k} g_{-k}(x)\hat{g}_k(x)\\
\hat{g}_B(x)&=&-\pin{p}e^{-ipx}\frac{\tilde{g}_B(p)}{2\omega_p}\hsp
\hat{g}_k(x)=\pin{p}e^{-ipx}\tilde{g}_k(p)\left(\frac{1}{2\ok{}}-\frac{1}{2\omega_p}\right).\nonumber
\eea
Using the completeness relations (\ref{pcomp}) one can show \cite{me2stato,wick} that $\I(x)$ is determined by
\beq
\partial_x \I(x)=\pin{k}\frac{1}{2\omega_k}\partial_x\left|g_{k}(x)\right|^2 \label{di}
\eeq
together with the condition that it vanish at spatial infinity.

\section{Translation Invariance} \label{transsez}

In this section we will calculate the translation operator that acts on our states $\mathcal{O}|\Omega\rangle$ and will use it to fix all $\phi_0$-descendants (components of states that include operators $\phi_0$).

\subsection{The Translation Operator}

Let us define the shorthand
\beq
\Delta_{ij}=\int dx g_i(x) g\p_j(x)=i\pin{p}p\tilde{g}_i(p)\tilde{g}_j(-p)
\eeq
where $i$ and $j$ may be a bound state or a momentum $k$.  Note that $\Delta$ is antisymmetric.  We will use reweighted creation and annihilation operators
\beq
B_k^\dag=\frac{b_k^\dag}{\sqrt{2\omega_k}}\hsp 
B_k=\sqrt{2\omega_k}b_k
\eeq
which satisfy the same Heisenberg commutation relations as $b^\dag$ and $b$.

The identity
\beq
P\df=\df\left(P-\sqrt{Q_0}\pi_0\right)
\eeq
implies that translation invariance
\beq
P|K\rangle=P\df\sum_i |0\rangle_i=0
\eeq
is equivalent to
\beq
P|0\rangle_i=\sqrt{Q_0}\pi_0|0\rangle_{i+1}. \label{ti}
\eeq
Our strategy will be to solve this equation by inverting $\pi_0$.  Thus translation invariance fixes our states entirely up to an element of the kernel of $\pi_0$.  We then {\it{only}} use the Schrodinger equation to fix the element of the kernel of $\pi_0$, thus greatly simplifying the problem.  Note that the kernel of $\pi_0$ consists precisely of the $\phi_0$-primary states.

Let us write the translation operator as
\bea
P&=&-\int dx \pi(x)\partial_x \phi(x)\\
&=&-\int dx\left[
\pi_0 g_B(x)\pin{k}\phi_k g\p_k(x)+\left(\pin{k} \pi_k g_k(x)\right)\phi_0 g\p_B(x)\right.\nonumber\\
&&\left.+\pink{2}\pi_{k_1}\phi_{k_2}g_{k_1}(x)g\p_{k_2}(x)
\right]\nonumber\\
&=&\pin{k}\Delta_{kB}\left[i\phi_0 \left(-\omega_kB_k^\dag+\frac{B_{-k}}{2}\right)+\pi_0\left(B_k^\dag+\frac{B_{-k}}{2\omega_k}\right)\right]\nonumber\\
&&+i\pink{2}\Delta_{k_1k_2}\left(-\omega_{k_1}B_{k_1}^\dag B_{k_2}^\dag+\frac{B_{-k_1}B_{-k_2}}{4\omega_{k_2}}-\frac{1}{2}\left(1+\frac{\omega_{k_1}}{\omega_{k_2}}\right)B^\dag_{k_1}B_{-k_2}
\right)\nonumber
\eea
and expand the $i$th order kink ground state as
\beq
\vac_i=\sum_{m,n=0}^\infty \vac_i^{mn}\hsp
\vac_i^{mn}=Q_0^{-i/2}\pink{n}\gamma_i^{mn}(k_1\cdots k_n)\phi_0^m\Bd1\cdots\Bd n\vac_0. \label{gameq}
\eeq
We will refer to $m=0$ states or matrix elements $\gamma_i^{0n}$ as $\phi_0$-primary and $m>0$ states as $\phi_0$-descendants.  Then translation invariance (\ref{ti}) yields the recursion relation
\bea
&&\gamma_{i+1}^{mn}(k_1\cdots k_n)=\left.\Delta_{k_n B}\left(\gamma_i^{m,n-1}(k_1\cdots k_{n-1})+\frac{\omega_{k_n}}{m}\gamma_i^{m-2,n-1}(k_1\cdots k_{n-1})\right)
\right. \label{rr}\\
&&+\pin{k\p}\Delta_{-k\p B}\sum_{j=0}^n\left(\frac{\gamma_i^{m,n+1}(k_1\cdots k_j,k\p,k_{j+1}\cdots k_n)}{2\omega_{k\p}}
-\frac{\gamma_i^{m-2,n+1}(k_1\cdots k_j,k\p,k_{j+1}\cdots k_n)}{2m}\right)\nonumber\\
&&+\frac{1}{2m}\sum_{j=1}^n\pin{k\p}\Delta_{k_n,-k\p}\left(1+\frac{\omega_{k_n}}{\omega_{k\p}}\right)\gamma^{m-1,n}_i(k_1\cdots k_{j-1}, k\p,k_j\cdots k_{n-1})
\nonumber\\
&&+\frac{\omega_{k_{n-1}}\Delta_{k_{n-1}k_n}}{m}\gamma_i^{m-1,n-2}(k_1\cdots k_{n-2})\nonumber\\
&&
\left.-\int\frac{d^2k\p}{(2\pi)^2}\frac{\Delta_{-k\p_1,-k\p_2}}{2m\omega_{k\p_2}}\sum_{j_1=1}^{n+1}\sum_{j_2=j_1+1}^{n+2} \gamma_i^{m-1,n+2}(k_1\cdots k_{j_1-1}, k\p_1, k_{j_1} \cdots k_{j_2-2},k\p_2,k_{j_2-1}\cdots k_{n}).
\right.
\nonumber
\eea
This recursion relation determines all $\phi_0$-descendants in terms of $\phi_0$-primary states plus the free state corresponding to the one-loop initial condition $\gamma_0$.   It does not determine the $\phi_0$-primaries, as it corresponds to a particular solution of Eq.~(\ref{ti}) and the addition of any element of the kernel of $\pi_0$, in other words any $\phi_0$-primary state, is another solution.


\subsection{Constructing Translation-Invariant States}

At  one loop, the quantum kink is described by a series of harmonic oscillators and so its spectrum is known precisely \cite{dhn1}.  To find a Hamiltonian eigenstate at higher but finite order, one need only start the recursion (\ref{rr}) at $i=0$ with the one-loop avatar of the state of interest.  

In this note we will apply this strategy to the ground state, corresponding to the initial condition
\beq
\gamma_0^{mn}=\delta_{m0}\delta_{n0}\gamma_0^{00}.
\eeq
One recursion yields
\beq
\gamma_1^{12}(k_1,k_2)=\omega_{k_1}\Delta_{k_1k_2}\gamma_0^{00}\hsp
\gamma_1^{21}(k_1)=\frac{\omega_{k_1}\Delta_{k_1B}}{2}\gamma_0^{00}. \label{g121}
\eeq
We are not interested in calculating the $\phi_0$-primaries ($m=0$ terms) because these are in the kernel of $\pi_0$ and so they are not determined by translation invariance.  These will be calculated using Schrodinger's equation in Sec.~\ref{ssez}.

We may continue by simply plugging in to our recursion relation (\ref{rr}).  But we can simplify things somewhat by noticing that (\ref{gameq}) does not completely determine the functions $\gamma_i^{mn}(k_1\cdots k_n)$.  For example, one may add any function which is antisymmetric under the exchange of any $k_i$ and $k_j$ without affecting $\vac$.  Therefore we are free to symmetrize each function.  As this will simplify our answer, that will be our convention: It will be understood that after calculating each $\gamma$ using (\ref{rr}) it should be symmetrized before the next recursion.  This convention allows one to perform all of the sums in our recursion relation (\ref{rr}), leaving
\bea
\gamma_{i+1}^{mn}(k_1\cdots k_n)&=&\left.\Delta_{k_n B}\left(\gamma_i^{m,n-1}(k_1\cdots k_{n-1})+\frac{\omega_{k_n}}{m}\gamma_i^{m-2,n-1}(k_1\cdots k_{n-1})\right)
\right. \label{rrs}\\
&&+(n+1)\pin{k\p}\Delta_{-k\p B}\left(\frac{\gamma_i^{m,n+1}(k_1\cdots k_n,k\p)}{2\omega_{k\p}}
-\frac{\gamma_i^{m-2,n+1}(k_1\cdots k_n,k\p)}{2m}\right)\nonumber\\
&&+\frac{\omega_{k_{n-1}}\Delta_{k_{n-1}k_n}}{m}\gamma_i^{m-1,n-2}(k_1\cdots k_{n-2})\nonumber\\
&&+\frac{n}{2m}\pin{k\p}\Delta_{k_n,-k\p}\left(1+\frac{\omega_{k_n}}{\omega_{k\p}}\right)\gamma^{m-1,n}_i(k_1\cdots k_{n-1},k\p)
\nonumber\\
&&\left.-\frac{(n+2)(n+1)}{2m}\int\frac{d^2k\p}{(2\pi)^2}\frac{\Delta_{-k\p_1,-k\p_2}}{2\omega_{k\p_2}} \gamma_i^{m-1,n+2}(k_1\cdots k_{n},k\p_1,k\p_2).
\right.
\nonumber
\eea

In summary, the recursion relation (\ref{rr}) always yields a correct $\gamma_{i+1}$ whereas the simpler (\ref{rrs}) is also correct if one first symmetrizes each $\gamma_i^{mn}(k_1\cdots k_n)$ with respect to its arguments $k_1\cdots k_n$.  Thus to apply (\ref{rrs}) to derive $\gamma_2$ we must first symmetrize all $\gamma_1^{mn}$ with $n\geq 2$.  We only found one such element, which after symmetrizing using the antisymmetry of $\Delta$ becomes
\beq
\gamma_1^{12}(k_1,k_2)=\frac{\left(\omega_{k_1}-\omega_{k_2}\right)\Delta_{k_1k_2}}{2}\gamma_0^{00}. \label{g112}
\eeq

What about the $\phi_0$-primaries $\gamma_1^{0n}$?  These are not fixed by translation invariance as they are in the kernel of $\pi_0$.  Rather they are determined using the Schrodinger equation.  In a scalar theory with a canonical kinetic term, $\phi$ will have dimensions of [action]${}^{1/2}$.  As each $\vac_{i}$ is suppressed by $\hbar^{1/2}$ with respect to $\vac_{i-1}$, it may only depend on terms in the potential up to $\phi^{2+i}$.  Therefore $\vac_1$ and so $\gamma_1$ only depend on $\phi^3$ terms.  As a result the only nonzero entries resulting from the Schrodinger equation can be $\gamma_1^{01}$ and $\gamma_1^{03}$.  

\begin{figure} 
\begin{center}
\includegraphics[width=2.5in,height=1.7in]{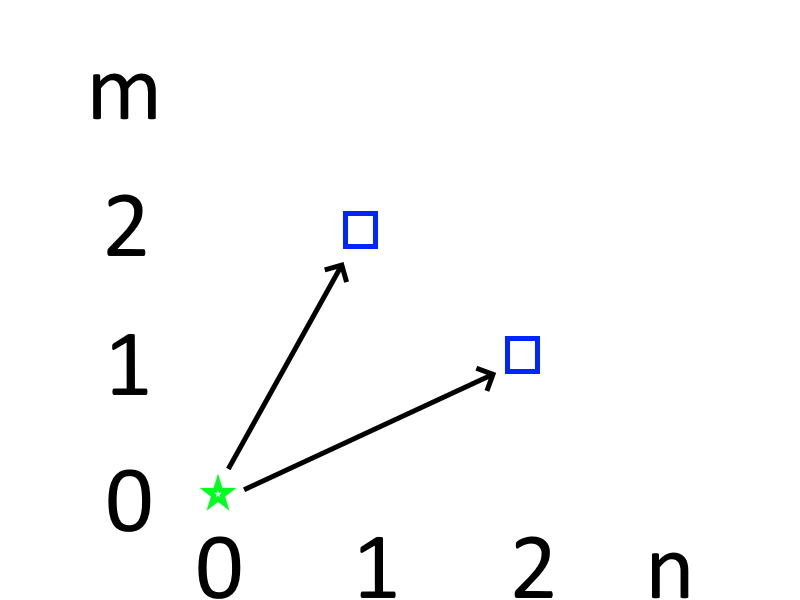}
\includegraphics[width=2.5in,height=1.7in]{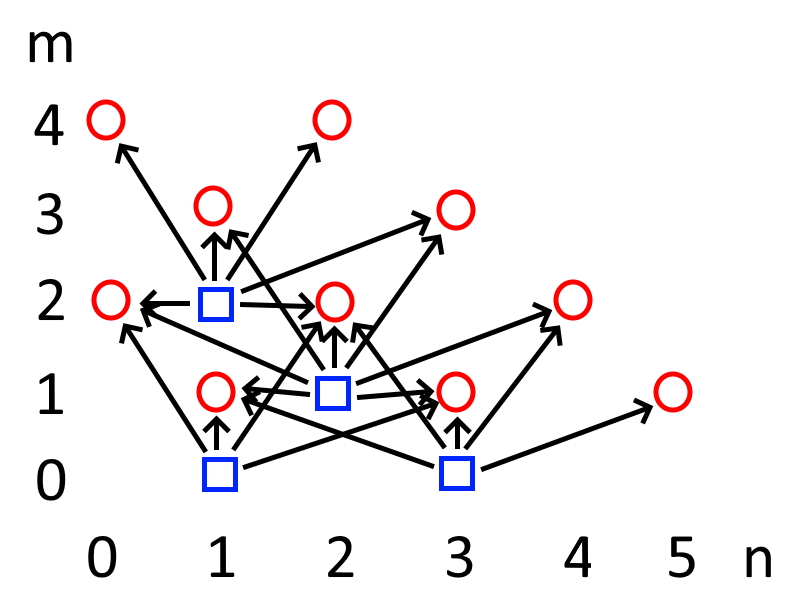}
\caption{The $\gamma_i^{mn}$ generated by the recursion relation at $i=1$ (left) and $i=2$ (right).  Green stars, blue squares and red circles represent elements at $i=0$, $i=1$ and $i=2$ respectively.  As $\phi_0$-primaries ($m=0$ elements) are in the kernel of $\pi_0$, they are not fixed by (\ref{ti}) and so arrows to such elements are not shown.}
\label{gfig}
\end{center}
\end{figure}

Finally we are ready to apply (\ref{rrs}) to calculate $\gamma_2$.  Remember that the recursion relations only determine $\phi_0$-descendants ($m>0$), so over all we expect 3, 4, 5 and 6 contributions from $\gamma_1^{01},\ \gamma_1^{03},\ \gamma_1^{21}$\ and $\gamma_1^{12}$ respectively.  At $m=1$ we find
\bea
\gamma_2^{11}(k_1)&=&\pin{k\p}\Delta_{-k\p B}\frac{\gamma_1^{12}(k_1,k\p)}{\omega_{k\p}}
-\frac{3}{4}\int\frac{d^2k\p}{(2\pi)^2}\frac{\Delta_{-k\p_1,-k\p_2}}{\omega_{k\p_2}} \gamma_1^{03}(k_1,k\p_1,k\p_2)\\
&&+\frac{1}{2}\pin{k\p}\Delta_{k_1,-k\p}\left(1+\frac{\omega_{k_1}}{\omega_{k\p}}\right)\gamma^{01}_1(k\p)\nonumber\\
&=&\frac{1}{2}\pin{k\p}\left(\frac{\omega_{k_1}}{\omega_{k\p}}-1\right)\Delta_{k_1k\p}\Delta_{-k\p B}\gamma_0^{00}
-\frac{3}{2}\int\frac{d^2k\p}{(2\pi)^2}\frac{\Delta_{-k\p_1,-k\p_2}}{\omega_{k\p_2}} \gamma_1^{03}(k_1,k\p_1,k\p_2)\nonumber\\
&&+\frac{1}{2}\pin{k\p}\Delta_{k_1,-k\p}\left(1+\frac{\omega_{k_1}}{\omega_{k\p}}\right)\gamma^{01}_1(k\p)\nonumber
\eea
and
\bea
\gamma_2^{13}(k_1,k_2,k_3)&=&\omega_{k_2}\Delta_{k_2k_3}\gamma_1^{01}(k_1)+\Delta_{k_3B}\gamma_1^{12}(k_1,k_2)\\
&&+\frac{3}{2}\pin{k\p}\Delta_{k_3,-k\p}\left(1+\frac{\omega_{k_3}}{\omega_{k\p}}\right)\gamma^{03}_1(k_1,k_2,k\p)
\nonumber\\
&=&\omega_{k_2}\Delta_{k_2k_3}\gamma_1^{01}(k_1)+\frac{1}{2}\Delta_{k_3B}\left(\omega_{k_1}-\omega_{k_2}\right)\Delta_{k_1k_2}\gamma_0^{00}\nonumber\\
&&+\frac{3}{2}\pin{k\p}\Delta_{k_3,-k\p}\left(1+\frac{\omega_{k_3}}{\omega_{k\p}}\right)\gamma^{03}_1(k_1,k_2,k\p)
\nonumber\\
\gamma_2^{15}(k_1\cdots k_5)&=&\omega_{k_4}\Delta_{k_4k_5}\gamma_1^{03}(k_1,k_2,k_3).\nonumber
\eea
Next at $m=2$
\bea
\gamma_2^{20}&=&\pin{k\p}\Delta_{-k\p B}\left(\frac{\gamma_1^{21}(k\p)}{2\omega_{k\p}}
-\frac{\gamma_1^{01}(k\p)}{4}\right)
-\frac{1}{4}\int\frac{d^2k\p}{(2\pi)^2}\frac{\Delta_{-k\p_1,-k\p_2}}{\omega_{k\p_2}} \gamma_1^{12}(k\p_1,k\p_2)\\
&=&\frac{1}{4}\pin{k\p}\Delta_{-k\p B}\left(\Delta_{k\p B}\gamma_0^{00}
-\gamma_1^{01}(k\p)\right)
+\frac{1}{8}\int\frac{d^2k\p}{(2\pi)^2}\left(1-\frac{\omega_{k\p_1}}{\omega_{k\p_2}}\right)\Delta_{k\p_1k\p_2}\Delta_{-k\p_1,-k\p_2}\gamma_0^{00}
\nonumber
\eea
and
\bea
\gamma_2^{22}(k_1,k_2)&=&\Delta_{k_2 B}\left(\gamma_1^{21}(k_1)+\frac{\omega_{k_2}}{2}\gamma_1^{01}(k_1)\right)-\frac{3}{4}\pin{k\p}\Delta_{-k\p B}
\gamma_1^{03}(k_1,k_2,k\p)\\
&&+
\frac{1}{2}\pin{k\p}\Delta_{k_2,-k\p}\left(1+\frac{\omega_{k_2}}{\omega_{k\p}}\right)\gamma^{12}_1(k_1,k\p)\nonumber\\
&=&\frac{\Delta_{k_2 B}}{2}\left(\omega_{k_1}\Delta_{k_1B}\gamma_0^{00}+\omega_{k_2}\gamma_1^{01}(k_1)\right)-\frac{3}{4}\pin{k\p}\Delta_{-k\p B}
\gamma_1^{03}(k_1,k_2,k\p)\nonumber\\
&&+
\frac{1}{4}\pin{k\p}\Delta_{k_2,-k\p}\left(1+\frac{\omega_{k_2}}{\omega_{k\p}}\right)
\left(\omega_{k_1}-\omega_{k\p}\right)\Delta_{k_1k\p}\gamma_0^{00}
\nonumber\\
\gamma_2^{24}(k_1\cdots k_4)&=&\frac{\omega_{k_{3}}\Delta_{k_{3}k_4}}{2}\gamma_1^{12}(k_1,k_{2})
+\Delta_{k_4 B}\frac{\omega_{k_4}}{2}\gamma_1^{03}(k_1\cdots k_3)
\nonumber\\
&=&\frac{\omega_{k_1}\omega_{k_{3}}\Delta_{k_1k_2}\Delta_{k_{3}k_4}}{2}\gamma_0^{00}
+\frac{\omega_{k_4}\Delta_{k_4 B}}{2}\gamma_1^{03}(k_1\cdots k_3)
.
\nonumber
\eea
Continuing to $m=3$ we find
\bea
\gamma_2^{31}(k_1)&=&-\frac{1}{3}\pin{k\p}\Delta_{-k\p B}
\gamma_1^{12}(k_1,k\p)
+\frac{1}{6}\pin{k\p}\Delta_{k_1,-k\p}\left(1+\frac{\omega_{k_1}}{\omega_{k\p}}\right)\gamma^{21}_1(k\p)
\nonumber\\
&=&\frac{\gamma_0^{00}}{6}\pin{k\p}\left[\left(\omega_{k\p}-\omega_{k_1}\right)\Delta_{k_1k\p}\Delta_{-k\p B}
+\frac{1}{2}\Delta_{k_1,-k\p}\left(\omega_{k_1}+\omega_{k\p}\right)
\omega_{k\p}\Delta_{k\p B}\right]
\nonumber\\
&=&\gamma_0^{00}\pin{k\p}\left(\frac{\omega_{k\p}}{4}-\frac{\omega_{k_1}}{12}\right)\Delta_{k_1k\p}\Delta_{-k\p B}
\nonumber\\
\gamma_2^{33}(k_1,k_2,k_3)&=&\frac{\omega_{k_3}\Delta_{k_3 B}}{3}\gamma_1^{12}(k_1,k_2)+\frac{\omega_{k_{2}}\Delta_{k_{2}k_3}}{3}\gamma_1^{21}(k_1)\nonumber\\
&=&\left(\omega_{k_3}\Delta_{k_3 B}\left(\omega_{k_1}-\omega_{k_2}\right)\Delta_{k_1k_2}+\omega_{k_{2}}\Delta_{k_{2}k_3}\omega_{k_1}\Delta_{k_1B}\right)\frac{\gamma_0^{00}}{6}.
\eea
Note that, since $\gamma_2^{33}$ is defined by its symmetric contraction with $\B1\B2\B3$, one is free to add any term which is annihilated by symmetrization of $k_1,\ k_2$\ and $k_3$.  Thus one may freely redefine
\beq
\gamma_2^{33}(k_1,k_2,k_3)=\frac{\omega_{k_1}\Delta_{k_1B}\omega_{k_{2}}\Delta_{k_{2}k_3}}{2}\gamma_0^{00}.
\eeq
In other words, different paths from $\gamma_0^{00}$ to $\gamma_2^{33}$ lead to contributions which are proportional.  This suggests that to some extent it may be possible to explicitly solve our recursion formula.  Finally the $m=4$ terms are
\bea
\gamma_2^{40}&=&-\pin{k\p}\Delta_{-k\p B}\frac{\gamma_1^{21}(k\p)}{8}
=\frac{\gamma_0^{00}}{16}\pin{k\p}
\omega_{k\p}\Delta_{B k\p}\Delta_{-k\p B}\\
\gamma_2^{42}(k_1,k_2)&=&\Delta_{k_2 B}\frac{\omega_{k_2}}{4}\gamma_1^{21}(k_1)=\frac{\omega_{k_1}\Delta_{k_1B}\omega_{k_2}\Delta_{k_2 B}}{8}\gamma_0^{00}.
\nonumber
\eea

\section{Schrodinger Equation} \label{ssez}

Let us define the symbol $\Gamma$ by any solution of
\beq
\sum_{j=0}^i \left(H_{i+2-j}-Q_{\frac{i-j}{2}+1}\right)\vac_j=Q_0^{-i/2}\sum_{mn} \pink{n} \Gamma_i^{mn}(k_1\cdots k_n)\phi_0^m B_{k_1}^\dag\cdots B_{k_n}^\dag\vac_0.\label{gdef}
\eeq
Then the Schrodinger Equation
\beq
(H-Q)\vac=0
\eeq
is solved if
\beq
\Gamma_i^{mn}=0.
\eeq
Note that $\Gamma$ is not uniquely defined by (\ref{gdef}).  A necessary and sufficient condition for a solution to Schrodinger's equations is that $\Gamma_i^{mn}$ vanishes when summed over all permutations of the $k_j$.  The number of loops can be defined by counting powers of $\hbar$ and is equal to $i/2+1$. Note that only integral numbers of loops correct the energy, and so $Q$ vanishes if its subscript is a half-integer.  Here $Q$ is defined to be the energy of the ground state.  For applications to other states, $Q$ should be replaced with their respective energies.

Let us begin with the one-loop approximation, $i=0$.  Using 
\beq
H_2-Q_1=\frac{\pi_0^2}{2}+\pin{k} \omega_k B_k^\dag B_k \label{libh}
\eeq
one finds that the Schrodinger equation is satisfied if
\beq
\pi_0\vac_0=B_k\vac_0=0.
\eeq
These are both satisfied by the initial condition  $\gamma_0^{mn}=\delta_{m0}\delta_{n0}$ of our recursion.

\subsection{Leading Corrections}

At $i=1$ the Schrodinger equation is
\beq
H_3\vac_0+(H_2-Q_1)\vac_1=0.
\eeq
Using
\beq
H_3=\frac{1}{6}\int dx \V3:\phi^3(x):_a=\frac{1}{6}\int dx \V3:\phi^3(x):_b+\frac{1}{2}\int dx \V3\phi(x) \I(x)
\eeq
where we have defined $\V{n}$ to be the $n$th derivative of $g^{-2}V[g\phi(x)]$ with respect to $\phi(x)$, evaluated at $\phi(x)=f(x)$, one finds that the leading correction to the states (\ref{g121}) yields
\bea
\Gamma_1^{21}&=&\sqrt{Q_0}\frac{V_{BBk_1}}{2}+\frac{\omega_{k_1}^2\Delta_{k_1B}}{2} \label{g1}\\
\Gamma_1^{12}&=&\sqrt{Q_0}\frac{V_{Bk_1k_2}}{2}+\frac{\left(\omega_{k_1}-\omega_{k_2}\right)\left(\omega_{k_1}+\omega_{k_2}\right)\Delta_{k_1B}}{2}\nonumber
\eea
where we have introduced the notation
\beq
V_{\I\stackrel{m}{\cdots}\I,\alpha_1\cdots\alpha_n}=\int dx V^{(2m+n)}[gf(x)]\I^m(x) g_{\alpha_1}(x)\cdots g_{\alpha_n(x)}
\eeq
where $\alpha_j$ can be $B$ or $k_j$. 

Substituting the identities \cite{me2stato}
\bea
 V_{BBk}&=&\int dx \V3 g_B(x)\frac{f\p(x)}{\sqrt{Q_0}}g_k(x)=\frac{1}{\sqrt{Q_0}}\int dx \partial_x\left(\V2\right)g_B(x)g_k(x)\nonumber\\
 &=&-\frac{1}{\sqrt{Q_0}}\int dx \V2\left(g\p_B(x)g_k(x)+g_B(x)g\p_k(x)\right)
 \nonumber\\
 &=&-\frac{1}{\sqrt{Q_0}}\int dx \left(g\p_B(x)\omega_k^2g_k(x)+g\p_B(x)g\pp_k(x)+g\pp_B(x)g\p_k(x)\right)
 \nonumber\\
 &=&-\frac{\omega_k^2}{\sqrt{Q_0}}\Delta_{kB}\nonumber\\
  V_{Bk_1k_2}&=&-\frac{1}{\sqrt{Q_0}}\int dx \V2\left(g\p_{k_1}(x)g_{k_2}(x)+g_{k_1}(x)g\p_{k_2}(x)\right)
 \nonumber\\
 &=&-\frac{1}{\sqrt{Q_0}}\int dx \left(g\p_{k_1}(x)\omega_{k_2}^2g_{k_2}(x)+g\p_{k_1}(x)g\pp_{k_2}(x)+\omega_{k_1}^2g_{k_1}(x)g_{k_2}\p(x)+g\pp_{k_1}(x)g\p_{k_2}(x) \right)
 \nonumber\\
 &=&\frac{\omega_{k_2}^2-\omega_{k_1}^2}{\sqrt{Q_0}}\Delta_{k_1 k_2} \label{vid}
 \eea
into (\ref{g1}) one finds $\Gamma=0$, and so these matrix elements of Schrodinger's equation are satisfied by the states (\ref{g121}), which were derived from translation invariance alone.  This is consistent with our claim that all $\phi_0$-descendants ($m>0$ components of states) are determined in terms of $\phi_0$-primaries by imposing the eigenvalue of the momentum, in this case zero.

The other components of the Schrodinger equation at $i=1$ are
\beq
\Gamma_1^{01}=\frac{\sqrt{Q_0}}{2}V_{\I k_1}-\frac{ \omega_{k_1}\Delta_{k_1B}}{2}+\omega_{k_1}\gamma_1^{01}\hsp
\Gamma_1^{03}=\frac{\sqrt{Q_0}}{6}V_{k_1k_2k_3}+(\omega_{k_1}+\omega_{k_2}+\omega_{k_3})\gamma_1^{03}
\eeq
and so the state at order $i=1$ is given by the $\phi_0$-descendants in Eqs.~(\ref{g121}) and (\ref{g112}) together with the $\phi_0$-primaries
\beq
\gamma_1^{01}=\frac{\Delta_{k_1B}}{2}-\frac{\sqrt{Q_0}}{2}Y_{\I k_1}\hsp
\gamma_1^{03}=-\frac{\sqrt{Q_0}}{6}Y_{k_1k_2k_3}
\eeq
where we have defined the reduced potential
\beq
Y_{k_1\cdots k_j}=\frac{V_{k_1\cdots k_j}}{\omega_{k_1}+\cdots+\omega_{k_j}}\hsp
Y_{\I,k_1\cdots k_j}=\frac{V_{\I,k_1\cdots k_j}}{\omega_{k_1}+\cdots+\omega_{k_j}}.
\eeq

Note that in models like the $\phi^4$ double well, in which the third derivative of the potential is nonzero at the minima, $V_{k_1k_2k_3}$ will have a divergence of the form $\delta\left(\sum_i k_i\right)$.  When integrated over $k$ to determine the state, this divergence leads to finite coefficients.  However at two loops it leads to an infrared divergence in the energy of the kink state.  As we will see in Subsec.~\ref{vacen}, this infrared divergence also appears in the vacuum energy and so the kink mass, which is the difference between the energies of the two states, is finite.

\begin{figure} 
\begin{center}
\includegraphics[width=2.5in,height=1.7in]{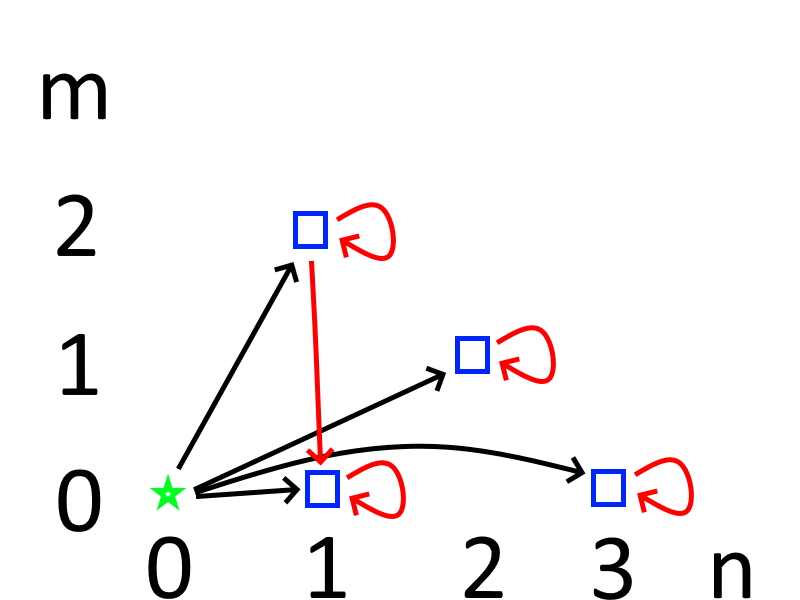}
\includegraphics[width=2.5in,height=1.7in]{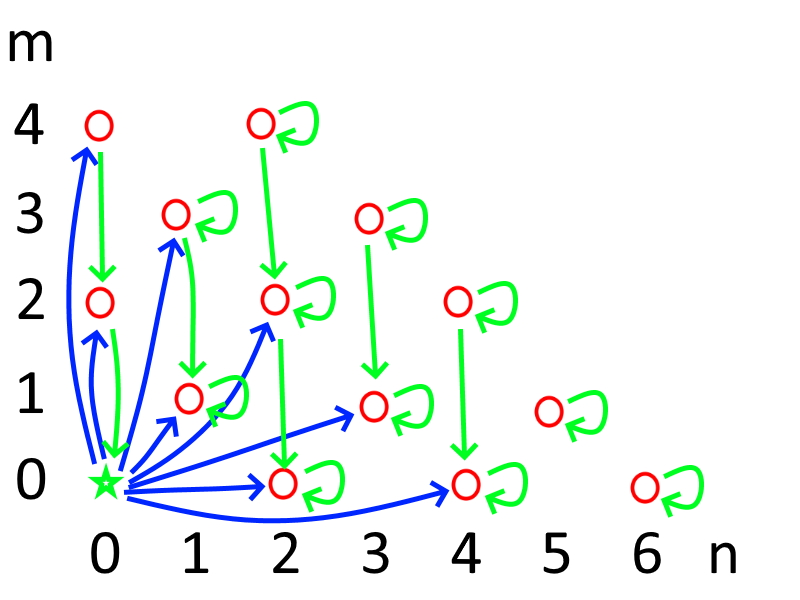}
\includegraphics[width=3.5in,height=2.5in]{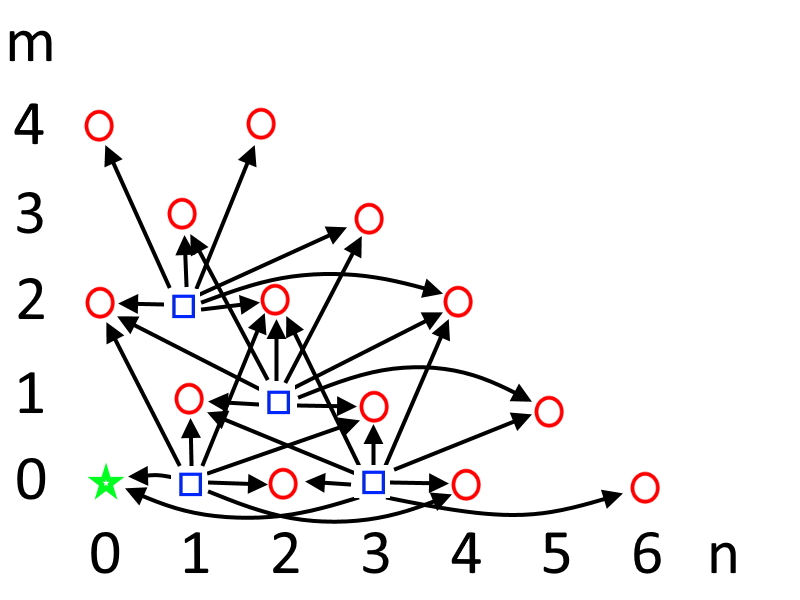}
\caption{Terms in the Schrodinger equation: (top left) $H_3\vac_0$ in black and $(H_2-Q_1)\vac_1$ in red, (top right) $(H_4-Q_2)\vac_0$ in blue and $(H_2-Q_1)\vac_2$ in green, (bottom) $H_3\vac_1$ in black.}
\label{hfig}
\end{center}
\end{figure}

\subsection{The Kink Ground State at Two Loops} \label{gamma0}

Translation invariance fixes all $\phi_0$-descendant components $\gamma_i^{mn}$ in any Hamiltonian eigenstate.  The $\phi_0$-primary terms $\gamma_i^{0n}$, at each order $i$ are fixed by the Schrodinger equation.   Interaction terms relate these coefficients to those at lower orders.  Thus the other $\gamma_i^{mn}$ are only related to $\gamma_i^{0n}$ by the free Hamiltonian (\ref{libh}).  More specifically, $\gamma_i^{0n}$ is related to $\gamma_i^{2n}$ by the $\pi_0^2/2$ term and to $\gamma_i^{0n}$ by the oscillator term.  This allows each $\phi_0$-primary $\gamma_i^{0n}$ to be determined from $\gamma_i^{2n}$ and the state at orders less than $i$.   In theories, like those considered here, with nonderivative interactions the situation is even simpler because interactions never decrease $m$.  Thus Schrodinger's equation determines $\phi_0$-primaries $\gamma_i^{0n}$ in terms of $\gamma_i^{2n}$ and $\phi_0$-primaries $\gamma_j^{0n\p}$ with $j<i$.  In other words, only the $\phi_0$-descendants at $m=2$ are needed.  Similarly the energy at each order $i$ is determined by $\gamma_i^{20}$ together with the $\phi_0$-primaries $\gamma_j^{0n}$ at lower orders $j<i$.   

This observation in practice leads to a dramatic reduction in the complexity of calculations of states and energies.  For example, to compute the two-loop energy of the kink ground state, one only needs to know $\gamma_2^{20}$, $\gamma_1^{01}$ and $\gamma_1^{03}$, which themselves are determined from $\gamma_1^{21}$ and $\gamma_1^{12}$.  In this subsection we will complete the calculation the kink ground state at two loops, corresponding to $i=2(2-1)=2$, by finding the $\phi_0$-primaries.

\subsubsection{$(m=0,n=6)$}

At $i=2$ the Schrodinger equation is
\beq
(H_4-Q_2)\vac_0+H_3\vac_1+(H_2-Q_1)\vac_2=0.
\eeq
Let us begin with the simplest element, $\Gamma_2^{06}$.  The previous argument agrees with Fig.~\ref{hfig} showing that there are two contributions, arising from $\gamma_1^{03}$, which was found at the previous order, and from $\gamma_2^{06}$ which is to be found now.  Defining the total energy
\beq
\Omega_n=\sum_{j=1}^n \omega_{k_j}
\eeq
these contributions are
\bea
H_3\vac_1^{03}&\supset& -\frac{1}{36}\pink{6}Y_{k_1k_2k_3}V_{k_4k_5k_6}B_{k_1}^\dag\cdots B_{k_6}^\dag\vac_0\\
H_2\vac_2^{06}&=&\frac{1}{Q_0}\pink{6}\Omega_6\gamma_2^{06}B_{k_1}^\dag\cdots B_{k_6}^\dag\vac_0
\nonumber
\eea
and so one finds the matrix element
\beq
\gamma_2^{06}=\frac{Q_0}{36}Y_{k_1k_2k_3}\frac{V_{k_4k_5k_6}}{\Omega_6}.
\eeq

\subsubsection{$(m=0,n=4)$}

To organize the calculations of the other matrix elements, we note that $\Gamma$ may be decomposed into contributions which do not mix with one another.  In particular contributions with different numbers of dummy momenta $k\p$ and with different numbers of powers of the undifferentiated\footnote{If multiplied by $V^{(4)}g_B(x)$ then an integration by parts leads to a differentiated $\I(x)$ which can be evaluated using (\ref{di}) and this argument does not apply.  This situation does not arise in the calculation of $\gamma_2^{0n}$ but does arise when verifying that the Schrodinger equation is satisfied in \ref{sapp}.} contraction factor $\I(x)$ together with $V^{(3)}$ do not mix.  We will include this decomposition in the subscript of $\Gamma$.  Of course each $\Gamma_i^{0n}$ determines $\gamma_i^{0n}$ whose form is not known before $\Gamma_i^{0n}$ is calculated, so terms resulting from $\gamma_i^{0n}$ will not be included in this decomposition.

Let us begin with all contributions $\Gamma_{2\I}^{04}$ containing a single power of the contraction factor $\I(x)$.  These contributions arise from two terms
\bea
H_3\vac_1^{03}&\supset& -\frac{1}{12}\pink{4}V_{\I k_4}Y_{k_1k_2k_3}B_{k_1}^\dag\cdots B_{k_4}^\dag\vac_0\\
H_3\vac_1^{01}&\supset& -\frac{1}{12}\pink{4}Y_{\I k_4}V_{k_1k_2k_3}B_{k_1}^\dag\cdots B_{k_4}^\dag\vac_0
\nonumber
\eea
whose sum yields
\beq
\Gamma_{2\I}^{04}=-\frac{Q_0}{12}Y_{\I k_4}Y_{k_1k_2k_3}\Omega_4.
\eeq

Next let us consider the contributions with one contracted momentum $k\p$.  There is only one
\beq
H_3\vac_1^{03}\supset -\frac{1}{8}\pink{4}\pin{k\p}Y_{k_1k_2-k\p}\frac{V_{k_3k_4-k\p}}{\omega_{k\p}}B_{k_1}^\dag\cdots B_{k_4}^\dag\vac_0
\eeq
yielding
\beq
\Gamma_{2k\p}^{04}= -\frac{Q_0}{8}\pin{k\p}Y_{k_1k_2-k\p}\frac{V_{k_3k_4-k\p}}{\omega_{k\p}}.
\eeq

Finally there are three contributions with no $k\p$
\bea
H_3\vac_1^{01}&\supset&\frac{1}{12\sqrt{Q_0}}\pink{4}\Delta_{k_1B}V_{k_2k_3k_4}B_{k_1}^\dag\cdots B_{k_4}^\dag\vac_0\\
\frac{\pi_0^2}{2}\vac_2^{24}&=&\frac{1}{12{Q_0}}\pink{4}\left[-6\omega_{k_1}\omega_{k_3}\Delta_{k_1k_2}\Delta_{k_3k_4}+\sqrt{Q_0}Y_{k_1k_2k_3}\omega_{k_4}\Delta_{k_4B}
\right]B_{k_1}^\dag\cdots B_{k_4}^\dag\vac_0\nonumber\\
H_4\vac_0&=&\frac{1}{24}\pink{4}V_{k_1k_2k_3k_4}B_{k_1}^\dag\cdots B_{k_4}^\dag\vac_0
\nonumber
\eea
which sum to
\beq
\Gamma_{2k^{\prime 0}}^{04}=\frac{\sqrt{Q_0}}{12}Y_{k_1k_2k_3}\Delta_{k_4 B}\Omega_4-\frac{\omega_{k_1}\omega_{k_3}}{2}\Delta_{k_1k_2}\Delta_{k_3k_4}+\frac{Q_0}{24}V_{k_1k_2k_3k_4}.
\eeq

The final contribution to $\Gamma_2^{04}$ arises from
\beq
\pin{k}\omega_k B_k^\dag B_{-k}\vac_{2}^{04}=\frac{1}{Q_0}\pink{4}\Omega_4\gamma_2^{04}B_{k_1}^\dag\cdots B_{k_4}^\dag\vac_0
\eeq
and is
\beq
\Gamma_{2f}^{04}=\Omega_4\gamma_2^{04}.
\eeq
The Schrodinger equation
\beq
0=\Gamma_2^{04}=\Gamma_{2f}^{04}+\Gamma_{2\I}^{04}+\Gamma_{2k^{\prime 0}}^{04}+\Gamma_{2k\p}^{04}
\eeq
then yields the matrix element
\bea
\gamma_2^{04}&=&-\frac{\Gamma_{2\I}^{04}+\Gamma_{2k\p}^{04}+\Gamma_{2k^{\prime 0}}^{04}}{\Omega_4}\\
&=&\frac{Q_0}{12}Y_{\I k_4}Y_{k_1k_2k_3}
-\frac{\sqrt{Q_0}}{12}Y_{k_1k_2k_3}\Delta_{k_4 B}+\frac{\omega_{k_1}\omega_{k_3}}{2\Omega_4}\Delta_{k_1k_2}\Delta_{k_3k_4}-\frac{Q_0}{24}Y_{k_1k_2k_3k_4}\nonumber\\
&&+\frac{Q_0}{8\Omega_4}\pin{k\p}Y_{k_1k_2-k\p}\frac{V_{k_3k_4-k\p}}{\omega_{k\p}}.
\nonumber
\eea

Note that in models like the Sine-Gordon model, in which the fourth derivative of the potential is nonzero at the minima, $Y_{k_1k_2k_3k_4}$ will have a divergence of the form $\delta\left(\sum_i k_i\right)$.  When integrated over $k$ to determine the state, this divergence leads to finite coefficients.  However at three loops it leads to an infrared divergence in the energy of the kink state.  As in the two-loop divergence in the $\phi^4$ kink energy, this divergence also appears in the vacuum energy and so the kink mass remains finite.  We expect such cancellations at all loops, as the infrared divergences arise from a regime in $x$ where $f(x)$ is equal to a vacuum value, and so the energy contribution from the kink and vacuum sector should agree.

\subsubsection{$(m=0,n=2)$}

The last matrix element needed to fix the ground state at two loops is $(m=0,n=2)$.  There is one contribution with two powers of the contraction factor $\I$
\beq
H_3\vac_1^{01}\supset -\frac{1}{4}\pink{2}Y_{\I k_1}V_{\I k_2} B_{k_1}^\dag B_{k_2}^\dag\vac_0
\eeq
which, after adding an antisymmetric term which does not affect the sum, yields
\beq
\Gamma_{2\I^2}^{02}=-\frac{Q_0}{8}Y_{\I k_1}Y_{\I k_2}\Omega_2.
\eeq
There are four contributions with a single power of $\I$
\bea
\frac{\pi_0^2}{2}\vac_2^{22}  &\supset& \frac{1}{4\sqrt{Q_0}}\pink{2} Y_{\I k_1}\omega_{k_2}\Delta_{k_2B}        B_{k_1}^\dag B_{k_2}^\dag\vac_0 
\\
H_3\vac_1^{01}  &\supset&\pink{2} \left( \frac{1}{4\sqrt{Q_0}}V_{\I k_2}\Delta_{k_1B}
-\frac{1}{8}\pin{k\p}Y_{\I k\p}\frac{V_{k_1k_2-k\p}}{\omega_{k\p}} 
    \right)   B_{k_1}^\dag B_{k_2}^\dag\vac_0 
\nonumber\\
H_4\vac_0  &\supset& \frac{1}{4} \pink{2} V_{\I k _1k_2}       B_{k_1}^\dag B_{k_2}^\dag\vac_0 
\nonumber\\
H_3\vac_1^{03}  &\supset& -\frac{1}{8} \pink{2} \pin{k\p}Y_{\I k\p}Y_{k_1k_2-k\p}        B_{k_1}^\dag B_{k_2}^\dag\vac_0 
\nonumber
\eea
which together contribute
\beq
\Gamma_{2\I}^{02}=\frac{\sqrt{Q_0}}{4}Y_{\I k_1}\Delta_{k_2 B}\Omega_2+\frac{Q_0V_{\I k_1k_2}}{4}-\frac{Q_0}{8}\pin{k\p}Y_{\I k\p}Y_{k_1k_2-k\p}\left(2+\frac{\Omega_2}{\omega_{k\p}}\right).
\eeq

Now we will organize the terms with no powers of $\I$ by the number of contracted momenta $k\p$.  There is one term with two contracted momenta
\beq
H_3\vac_1^{03}  \supset -\frac{1}{8}\pink{2}\pinkp{2}Y_{k_1k\p_1k\p_2} \frac{V_{k_2-k\p_1-k\p_2}}{\omega_{k\p_1}\omega_{k\p_2}}   B_{k_1}^\dag B_{k_2}^\dag\vac_0 
\eeq
yielding
\beq
\Gamma_{2k^{\prime 2}}^{02}=-\frac{Q_0}{8}\pinkp{2}Y_{k_1k\p_1k\p_2} \frac{V_{k_2-k\p_1-k\p_2}}{\omega_{k\p_1}\omega_{k\p_2}}.
\eeq
There are two sources of terms with no $\I$ and a single $k\p$
\bea
H_3\vac_1^{01}  &\supset& \frac{1}{8\sqrt{Q_0}}\pink{2} \pin{k\p}\frac{V_{k_1k_2k\p}}{\omega_{k\p}}\Delta_{-k\p B}   B_{k_1}^\dag B_{k_2}^\dag\vac_0 \\
\frac{\pi_0^2}{2}\vac_2^{22}  &\supset& \pink{2}\pin{k\p}\left[
-\frac{1}{8\sqrt{Q_0}}Y_{k_1k_2k\p}\Delta_{-k\p B}\right.\nonumber\\
&&\left.+\frac{1}{4Q_0}\Delta_{k_1k\p}\Delta_{-k\p k_2}\left(1+\frac{\omega_{k_2}}{\omega_{k\p}}\right)\left(\omega_{k_1}-\omega_{k\p}\right)
\right]    B_{k_1}^\dag B_{k_2}^\dag\vac_0 
\nonumber
\eea
which contribute
\beq
\Gamma_{2k^{\prime}}^{02}=\pin{k\p}\left[
\frac{\sqrt{Q_0}}{8} \frac{\Omega_2}{\omega_{k\p}}Y_{k_1k_2k\p}\Delta_{-k\p B}+\frac{1}{4}\Delta_{k_1k\p}\Delta_{-k\p k_2}\left(\frac{\omega_{k_1}\omega_{k_2}}{\omega_{k\p}}-\omega_{k\p}\right)
\right].
\eeq
Finally the terms with neither $\I$ nor $k\p$ are
\beq
\frac{\pi_0^2}{2}\vac_2^{22}\supset -\frac{3}{8Q_0}   \pink{2} \Omega_2 \Delta_{k_1 B}\Delta_{k_2B}    B_{k_1}^\dag B_{k_2}^\dag\vac_0 
\eeq
and so
\beq
\Gamma_{2k^{\prime 0}}^{02}=-\frac{3}{8} \Omega_2 \Delta_{k_1 B}\Delta_{k_2B}.
\eeq

As in the previous cases,
\beq
\Gamma_{2f}^{02}=\Omega_2\gamma_2^{02}
\eeq
and so the Schrodinger equation
\beq
0=\Gamma_2^{02}=\Gamma_{2f}^{02}+\Gamma_{2\I^2}^{02}+\Gamma_{2\I}^{02}+\Gamma_{2k^{\prime 2}}^{02}+\Gamma_{2k\p}^{02}+\Gamma_{2k^{\prime 0}}^{02}
\eeq
fixes the last matrix element
\bea
\gamma_2^{02}&=&-\frac{\Gamma_{2\I^2}^{02}+\Gamma_{2\I}^{02}+\Gamma_{2k^{\prime 2}}^{02}+\Gamma_{2k\p}^{02}+\Gamma_{2k^{\prime 0}}^{02}}{\Omega_2}\\
&=&\frac{Q_0}{8}Y_{\I k_1}Y_{\I k_2}-\frac{\sqrt{Q_0}}{4}Y_{\I k_1}\Delta_{k_2 B}-\frac{Q_0V_{\I k_1k_2}}{4\Omega_2}
+\frac{3}{8} \Delta_{k_1 B}\Delta_{k_2B}
\nonumber\\
&+&\frac{Q_0}{8\Omega_2}\pinkp{2}Y_{k_1k\p_1k\p_2} \frac{V_{k_2-k\p_1-k\p_2}}{\omega_{k\p_1}\omega_{k\p_2}}+\pin{k\p}\left[-
\frac{\sqrt{Q_0}}{8} \frac{1}{\omega_{k\p}}Y_{k_1k_2k\p}\Delta_{-k\p B}\right.\nonumber\\
&+&\left.\frac{1}{4\Omega_2}\left(
Q_0Y_{\I k\p}Y_{k_1k_2-k\p}\left(1+\frac{\Omega_2}{2\omega_{k\p}}\right)+
\Delta_{k_1k\p}\Delta_{-k\p k_2}\left(\omega_{k\p}-\frac{\omega_{k_1}\omega_{k_2}}{\omega_{k\p}}\right)\right)
\right].
\nonumber
\eea

\section{The Kink Mass} \label{msez}


\subsection{The Energy of the Kink Ground State}

The last Schrodinger equation is $\Gamma_2^{00}=0$.  This does not fix $\gamma_2^{00}$ because $\Gamma_2^{00}$ does not depend on $\gamma_2^{00}$.  This is reasonable because any value of $\gamma_2^{00}$ can be absorbed into the normalization of the state.  Thus one may normalize the ground state so that
\beq
\gamma_i^{00}=\delta_{i0}.
\eeq

Let us now solve this last Schrodinger equation.  There are two terms with two powers of the contraction factor $\I$
\bea
H_4\vac_0&\supset&\frac{V_{\I\I}}{8}\vac_0\\
H_3\vac_1^{01}&\supset&-\frac{1}{8}\pin{k\p}Y_{\I k\p}Y_{\I -k\p}\vac_0\nonumber
\eea
yielding
\beq
\Gamma_{2\I^2}^{00}\supset\frac{Q_0}{8}\left(V_{\I\I}-\pin{k\p}Y_{\I k\p}Y_{\I -k\p}\right).
\eeq
There are also two terms with a single factor of $\I$
\bea
H_3\vac_1^{01}&\supset&\frac{1}{8\sqrt{Q_0}}\pin{k\p}Y_{\I k\p}\Delta_{-k\p B}\vac_0\\
\frac{\pi_0^2}{2}\vac_2^{20}&\supset&-\frac{1}{8\sqrt{Q_0}}\pin{k\p}Y_{\I k\p}\Delta_{-k\p B}\vac_0\nonumber
\eea
which precisely cancel.
The terms with no factors of $\I$ can be organized by the number of contracted momenta $k\p$.  There is one term with 3, 2 and 1 momenta respectively, which for brevity we summarize together
\bea
H_3\vac_1^{03}&\supset&-\frac{1}{48}\pinkp{3}Y_{k_1\p k_2\p k_3 \p} \frac{V_{-k_1\p -k_2\p -k_3\p}}{\okp1\okp2\okp3}\\
\frac{\pi_0^2}{2}\vac_2^{20}&\supset&\frac{1}{16Q_0}\pinkp{2}\frac{\left(\omega_{k_1\p}-\omega_{k_2\p}\right)^2}{\omega_{k\p_1}\omega_{k\p_2}}\Delta_{k_1\p k_2\p}\Delta_{-k_1\p -k_2\p}\nonumber\\
\frac{\pi_0^2}{2}\vac_2^{20}&\supset&-\frac{1}{8Q_0}\pin{k\p}\Delta_{B k\p}\Delta_{B -k\p}.
\nonumber
\eea
As
\beq
g_k^*(x)=g_{-k}(x)
\eeq
the symbols $\Delta$, $V$ and $Y$ are all complex conjugated when all of their $k$ arguments are negated.  Therefore these contributions can each be rewritten as norms squared and so are real.  The corresponding $\Gamma$ can therefore be written
\bea
\Gamma_{2k^{\prime 3}}^{00}&=&   -\frac{Q_0}{48}\pinkp{3} \frac{\left|V_{k\p_1k\p_2k\p_3}\right|^2}{\omega_{k\p_1}\omega_{k\p_2}\omega_{k\p_3}\left(\omega_{k\p_1}+\omega_{k\p_2}+\omega_{k\p_3}\right)}  \\
\Gamma_{2k^{\prime 2}}^{00}&=&   \frac{1}{16}\pinkp{2}\frac{\left|\left(\omega_{k_1\p}-\omega_{k_2\p}\right)\Delta_{k_1\p k_2\p}\right|^2}{\omega_{k\p_1}\omega_{k\p_2}}
   \nonumber\\
\Gamma_{2k^{\prime}}^{00}&=&     -\frac{1}{8}\pin{k\p}\left|\Delta_{k\p B}\right|^2 . 
\nonumber
\eea

The last term may be written in a more convenient form using the completeness relation (\ref{comp})
\bea
\pin{k}\Delta_{kB}\Delta_{-kB}&=&\frac{1}{Q_0}\int dx\int dy \pin{k} g_{k}(x)g_{-k}(y)f\pp(x)f\pp(y)\\
&=&\frac{1}{Q_0}\int dx\int dy \left(\delta(x-y)-g_B(x)g_B(y)\right)f^{\prime\prime}(x)f^{\prime\prime}(y)=\frac{1}{Q_0}\int dx |f^{\prime\prime}(x)|^2\nonumber
\eea
where the $g_B(x)f^{\prime\prime}(x)$ integrals vanish because they are proportional to the total derivative of $g_B^2(x)$.

The Schrodinger equation then gives the two-loop energy
\bea
Q_2&=&\frac{1}{Q_0}\left(\Gamma_{2\I^2}^{00}+\Gamma_{2k^{\prime 3}}^{00}+\Gamma_{2k^{\prime 2}}^{00}+\Gamma_{2k^{\prime }}^{00}\right) \label{q2}\\
&=&\frac{V_{\I\I}}{8}-\frac{1}{8}\pin{k\p}\left|Y_{\I k\p}\right|^2
-\frac{1}{48}\pinkp{3} \frac{\left|V_{k\p_1k\p_2k\p_3}\right|^2}{\omega_{k\p_1}\omega_{k\p_2}\omega_{k\p_3}\left(\omega_{k\p_1}+\omega_{k\p_2}+\omega_{k\p_3}\right)}\nonumber\\
&&  +\frac{1}{16Q_0}\pinkp{2}\frac{\left|\left(\omega_{k_1\p}-\omega_{k_2\p}\right)\Delta_{k_1\p k_2\p}\right|^2}{\omega_{k\p_1}\omega_{k\p_2}}  -\frac{1}{8Q_0}\pin{k\p}
\left|f^{\prime\prime}(x)\right|^2 . 
\nonumber
\eea

To our knowledge, this is the first time that the two-loop energy has been calculated for kinks that need be neither integrable nor supersymmetric.  The explicit calculation of Refs.~\cite{vega,verwaest}, in the case of the Sine-Gordon model, did not require integrability and so could be repeated in this general setting.  However in that case we stress that the energy was found by summing 13 divergent Feynman diagrams, and carefully regulating and subtracting the divergences.  Here instead we find 5 terms, each of which is already UV finite.  Let us identify each of these terms.

When changing from plane wave to normal mode normal ordering, so that $H_2$ annihilates the one-loop kink ground state, the interaction Hamiltonian acquired constant and tadpole terms.  The first two terms in (\ref{q2}) are just the corresponding leading shifts to the energy, equal to the constant $V_{\I\I}$ plus the first perturbative contribution from the tadpole $V_{\I k\p}$.  The next term is the usual one-loop perturbation theory correction to an energy arising from a cubic interaction, and is given by the same expression as in the vacuum sector (\ref{vaccorr}) with plane waves replaced by normal modes.  The fourth term is a correction to the third term arising from the fact that derivative operators mix the normal modes, which is not the case for plane waves.  The last term was found long ago \cite{gjs,vega} using the collective coordinate approach, where it appeared as the leading term in an expansion of the denominator of an effective Hamiltonian, which came from a canonical transformation that separated a nonlinear extension of $\phi_B(x)$.  The manipulations which led to its appearance here are very different from those in the collective coordinate approach, but it is reassuring to see this agreement in the result.

The only trace of renormalization can be found in the first two terms, in the function $\I(x)$ which is the expected difference between two divergent sums weighted by $1/\omega_k$ and $1/\omega_p$ respectively.  In models such as the $\phi^4$ double well, in which the potential has a nonvanishing third derivative at the minima, the third term will be IR divergent.  This divergence arises from the region far from the kink, and so its contribution to the kink mass will be canceled by the same IR divergence in the vacuum energy, which we will now calculate.


\subsection{Vacuum Sector Energy} \label{vacen}

The kink mass is generally not $Q_2$.  It is $Q_2-E_1$ where $E_1$ is the 1-loop correction to the vacuum sector energy, as this contributes at the same order.  It is easily computed in perturbation theory.  Decompose the field in terms of plane waves as 
\beq
\phi(x)=\pin{p}\left(A_p^\dag+\frac{A_{-p}}{2\omega_p}\right)e^{-ipx}
\eeq
and the free and interaction Hamiltonians can be written
\beq
H_2=\pin{p} \omega_p A^\dag_p A_p\hsp
H_{n>2}= \frac{1}{n!} \int dx V^{(n)}[\phi_0]:\phi^n(x):_a
\eeq
where $\phi_0$ is the minimum of $V$ corresponding to the vacuum.   Then the first order of perturbation theory
\beq
H_3|\Omega\rangle_0+H_2|\Omega\rangle_1=0
\eeq
yields the first order correction $|\Omega\rangle_1$ to the vacuum state $|\Omega\rangle$
\beq
|\Omega\rangle_1=-\frac{V^{(3)}[\phi_0]}{6}\int\frac{d^3 p}{(2\pi)^3}\frac{2\pi\delta(p_1+p_2+p_3)}{\omega_{p_1}+\omega_{p_2}+\omega_{p_3}} A_{p_1}^\dag A_{p_2}^\dag A_{p_3}^\dag|\Omega\rangle_0.
\eeq
Acting again with $H_3$, the $|\Omega\rangle_0$ term yields the one loop correction to the energy
\bea
H_3|\Omega\rangle_1&\supset& -\frac{\left(V^{(3)}[\phi_0]\right)^2}{48}\int dx \pinpp{3} e^{-ix(p_1\p+p_2\p+p_3\p)}\frac{2\pi\delta(p_1+p_2+p_3)}{\omega_{p_1\p}\omega_{p_2\p}\omega_{p_3\p}\left(\omega_{p_1\p}+\omega_{p_2\p}+\omega_{p_3\p}\right)}|\Omega\rangle_0\nonumber\\
&=&-\frac{\left(V^{(3)}[\phi_0]\right)^2}{48} L \pinpp{3}\frac{2\pi\delta(p_1+p_2+p_3)}{\omega_{p_1\p}\omega_{p_2\p}\omega_{p_3\p}\left(\omega_{p_1\p}+\omega_{p_2\p}+\omega_{p_3\p}\right)}|\Omega\rangle_0
 \label{h31}
\eea
where $L$ is the length of the spatial direction\footnote{Here we are cavalier with boundary conditions, as the theory contains only scalar fields.  In practice, we simply subtract the kink and vacuum energy densities before performing the $x$ integration, in which case the integral converges.  In a theory with fermions a more careful approach may be warranted, for example adding a distant antikink to each kink to allow identical boundary conditions in each sector.}  which serves as an infrared cut off.   

The subleading correction to the Schrodinger equation is
\beq
(H_4-E_1)|\Omega\rangle_0+H_3|\Omega\rangle_1+H_2|\Omega\rangle_2=0.
\eeq
As $H_4$ is normal ordered, $H_4|\Omega\rangle_0$ is orthogonal to $|\Omega\rangle_0$ and so does not contribute to $E_1$.  We will chose $|\Omega\rangle_2$ to be orthogonal to $|\Omega\rangle_0$ so that the last term does not contribute to $E_1$.   Then $E_1$ can be read off of (\ref{h31}).  Evaluating the delta function, this is
\beq
E_1=-\frac{\left(V^{(3)}[\phi_0]\right)^2}{48} L \pinpp{2}\frac{1}{\omega_{p_1\p}\omega_{p_2\p}\omega_{p_1\p+p_2\p}\left(\omega_{p_1\p}+\omega_{p_2\p}+\omega_{p_1\p+p_2\p}\right)}. \label{vaccorr}
\eeq
The dependence on the infrared cutoff $L$ implies that we have calculated an energy density, and not an energy.  When this energy density is nonvanishing, it must be subtracted from the kink ground state energy to obtain the kink mass.  The kink mass will be finite only if these divergences cancel.  This procedure depends on the matching of the infrared divergences, which can be achieved for example if the energy densities are subtracted before they are integrated.  If the potential is symmetric about the minimum $\phi_0$, as it is in the case of the Sine-Gordon model but not the $\phi^4$ double well, $V^{(3)}[\phi_0]$ vanishes and so $E_1=0$ and this complication is avoided.

\subsection{The Sine-Gordon Model}

In the case of the Sine-Gordon model, the two-loop mass has been conjectured in \cite{dhn2} and calculated in \cite{vega,verwaest,nastase,zamkink}.  It is of course dependent upon the renormalization scheme \cite{dhn2} although in some schemes there is a renormalization group flow invariant coupling which provides a universal relation between the kink and meson mass.  No such relation may be expected to hold in general as there are other schemes in which the coupling may be shifted by any finite amount at any scale.

Using the well-known Sine-Gordon normal modes \cite{me2stato}
\beq
g_k(x)=\frac{e^{-ikx}{\rm{sign}}(k)}{\sqrt{1+m^2/k^2}}\left(k-i m\tanh(m x)\right)\hsp
g_{B}(x)=\sqrt{\frac{m}{2}}\sech\left(mx\right).  \label{geq}
\eeq
a contour integration yields
\bea
\Delta_{kB}&=&\frac{i\pi\ok{}}{\sqrt{8M}}{\rm{sech}}\left(\frac{k\pi}{2M}\right)\label{dkk}\\
\Delta_{k_1k_2}&=&-i(k_1-k_2)\pi\delta(k_1+k_2)+\frac{i\pi}{2}\frac{(k_2^2-k_1^2)}{\omega_{k_1}\omega_{k_2}}\rm{csch}\left(\frac{\pi\left(k_1+k_2\right)}{2m}\right)\rm{sign}(k_1k_2). \nonumber
\eea
We have evaluated the energy (\ref{q2}) term by term.  

In the first two terms, $\I(x)$ appears.  It was calculated in Ref.~\cite{me2stato} by integrating the general identity (\ref{di}).  Using the present conventions
\beq
\I(x)=-\frac{\rm{sech}^2(Mx)}{2\pi}.
\eeq
This leads to $Mg^2/(40\pi^2)$ and $-Mg^2/(120\pi^2)$ for the first and second terms of (\ref{q2}).  In the fourth term, the delta function in (\ref{dkk}) is multiplied by a zero which leads to a vanishing contribution, as can be checked directly by considering the case $k_1=k_2$ separately from the beginning.  Fixing the mass $M$ and coupling $g$ to unity, the third, fourth and fifth terms are equal to terms which may be found in Ref.~\cite{vega} and they were evaluated analytically by Verwaest who found that the sum of the third and fourth is $-Mg^2/(60\pi^2)$ while the fifth is $-Mg^2/192$.   Altogether we find that the two-loop correction to the kink mass is
\beq
Q_2=-\frac{Mg^2}{192}
\eeq
in agreement with the literature.  As shown in \ref{mapp} our normal ordering prescription yields the same meson mass as Ref.~\cite{vega} and so the soliton to meson mass ratio agrees, in accordance with Refs.~\cite{dhn2,zamkink}.



\section{Remarks}

Calculations of masses of quantum kinks have been an industry from Ref.~\cite{dhn1} to Refs.~\cite{aguirre,takyi}.  So far these calculations have been largely at one loop, where they are described by a free theory, with the exception of integrable and supersymmetric models.  In this paper, we have calculated the two-loop masses of scalar kinks in theories with arbitrary potentials.  We have also explicitly constructed their states, with the $\phi_0$-descendants calculated in Sec.~\ref{transsez} using translation invariance and the $\phi_0$-primary components in Sec.~\ref{ssez} using the Schrodinger equation.  These constructions we feel are even more interesting than the masses, as they allow one to compute matrix elements and so open the door to understanding the phenomenology \cite{kinkfen}, such as scattering \cite{adamscat,chris,wobble} and acceleration \cite{melac,melac2} of quantum kinks beyond the harmonic oscillator approximation.  For example, one may calculate form factors \cite{royff,kimform}.

While we only calculated the ground state, starting our recursion with a superposition of normal modes would have allowed us to apply the same strategy to an arbitrary state in the one-kink sector. 

The key step in our calculation was perturbatively imposing the translation invariance conditions, which fixed most matrix elements of the state, the $\phi_0$-descendents, in terms of a few coefficients, the $\phi_0$-primaries.  The $\phi_0$-primary components needed to be fixed using ordinary perturbation theory.  More generally, in the case of any translation-invariant Hamiltonian, as the Hamiltonian and momentum operators commute, a basis of all Hamiltonian eigenstates may be obtained by first fixing the momentum to obtain the $\phi_0$-descendant matrix elements in terms of $\phi_0$-primary matrix elements $\gamma_i^{0n}$ and then using the Schrodinger equation to fix the $\phi_0$-primary matrix elements.

In the case of a BPS state in a supersymmetric model one may first impose both translation invariance and also that the state be invariant under the preserved supersymmetries.  Presumably this will strongly constrain the state.  The big question is whether in a sufficiently supersymmetric model, this may constrain the state sufficiently that perturbation theory is no longer required.  In this case, one would have finally opened the door to a truly quantum understanding of nonperturbative solitons.  More precisely, one could understand the physical mechanisms at work behind the nonrenormalization theorems.  This of course is a prerequisite for applying lessons from supersymmetric theories to Yang-Mills.

\appendix
\section{Checking Schrodinger's Equation} \label{sapp}

We have derived the two-loop ground state using translation invariance together with Schrodinger's equation.  We restricted Schrodinger's equation to the $\phi_0$-primaries, the subspace of the Fock space with no $\phi_0$ acting on $\vac_0$, but we argued that, since the Hamiltonian and momentum operators commute, we expect our solutions to solve the Schrodinger equation in the full Fock space.  By imposing a condition on the momentum it is not possible that we lose the ground state solution, since it indeed must have zero momentum.  Furthermore, since the solution that we find, given the one-loop contribution, is unique, it must be the ground state.

In this Appendix we explicitly check this claim by inserting our two-loop state into the Schrodinger equation and showing that it vanishes on the full Fock space.  More precisely, we compute the various $\phi_0$-descendant components $\Gamma_2^{mn}$ at $m>0$ and show that they each vanish as claimed.  Recall that in Subsec.~\ref{gamma0} we found the $\phi_0$-primaries $\gamma_2^{0n}$ by imposing that $\Gamma_2^{0n}$ vanishes, and so we already know that the $m=0$ Schrodinger equation is satisfied.

\subsubsection{$m=5,\ n=1$}
The only contribution
\beq
H_3\vac_1^{21}\supset \frac{1}{6}\int dx \vt g_B^3(x)\phi_0^3\vac_1^{21}=0
\eeq
vanishes because 
\bea
V_{BBB}&=&\int dx \V3 g_B^3(x)=\frac{1}{\sqrt{Q_0}}\int dx \left(\partial_x\V2\right)g_B^2(x)\\
&=&-\frac{2}{\sqrt{Q_0}}\int dx\V2 g_B(x) g_B\p(x)=-\frac{2}{\sqrt{Q_0}}\int dx g\pp_B(x) g_B\p(x)=0\nonumber
\eea
is a total derivative.
 
 \subsubsection{$m=4,\ n=2$}
 \beq
 H_3\vac_1^{21}\supset \frac{1}{4\sqrt{Q_0}}\pink{2}\omega_{k_1}\Delta_{k_1B}V_{k_2 BB}\phi_0^4\B1\B2\vac_0
 \eeq
 exactly cancels
 \beq
 \pin{k\p}\okp{}\Bp{}B_{k\p}\vac_2^{42}=\frac{1}{4Q_0}\pink{2}\ok{1}^2\ok{2}\Delta_{k_1B}\Delta_{k_2B}\phi_0^4\B1\B2\vac_0
 \eeq
 as a result of (\ref{vid}).
 
 \subsubsection{$m=4,\ n=0$}
 \beq
 H_3\vac_1^{21}\supset -\frac{1}{8}\pink{1}Y_{kBB}Y_{-kBB}\phi_0^4\vac_0
 \eeq
 exactly cancels
 \beq
 H_4\vac_0\supset \frac{V_{BBBB}}{24}\phi_0^4\vac_0
 \eeq
 as
 \bea
 V_{BBBB}&=&\int dx \V4 g^4_B(x)=\int dx \V4 g^3_B(x)\frac{f\p(x)}{\sqrt{Q_0}}\label{vbbbb}\\
 &=&\frac{1}{\sqrt{Q_0}}\int dx \partial_x\left(\V3\right)g_B^3(x)=-\frac{3}{\sqrt{Q_0}}\int dx\V3 g_B^2(x)g\p_B(x)\nonumber\\
 &=&-\frac{3}{\sqrt{Q_0}}\int dx\V3 g_B^2(x)\int dy \delta(x-y)g\p_B(y)\nonumber\\
  &=&-\frac{3}{\sqrt{Q_0}}\int dx\V3 g_B^2(x)\int dy 
 \left[ g_B(x)g_B(y)+\pin{k}g_k(x)g_{-k}(y)\right]
  g\p_B(y)\nonumber\\
  &=&-\frac{3}{\sqrt{Q_0}}\pin{k}V_{kBB}\Delta_{-kB}=3\pin{k}Y_{kBB}Y_{-kBB}.\nonumber
 \eea
 
\subsubsection{$m=3,\ n=3$}
The two terms in
 \bea
  &&\pin{k\p}\okp{}\Bp{}B_{k\p}\vac_2^{33}\\
  &&=-\frac{1}{4\sqrt{Q_0}}\pink{3}\left[V_{k_1k_2B}\ok3\Delta_{k_3B}+V_{k_3BB}\left(\ok1-\ok2\right)\Delta_{k_1k_2}\right]\phi_0^3\B1\B2\B3\vac_0\nonumber
  \eea
 are respectively canceled by $H_3\vac_1^{21}$ and $H_3\vac_1^{12}$.

\subsubsection{$m=3,\ n=1$}
We will need the identity
\bea
V_{\I B}&=&\int dx \V3 g_B(x)\I(x)=\frac{1}{\sqrt{Q_0}}\int dx \left(\partial_x\V2\right)\I(x)\\
&=&-\frac{1}{\sqrt{Q_0}}\int dx\V2 \I\p(x)=-\frac{1}{\sqrt{Q_0}}\int dx\V2\pin{k}\frac{g_k(x)g\p_{-k}(x)}{\omega_k}\nonumber\\
&=&-\frac{1}{\sqrt{Q_0}}\pin{k}\int dx\left(\omega_k g_k(x)+\frac{g\pp_k(x)}{\omega_k}\right)g\p_{-k}(x)\nonumber\\
&=&-\frac{1}{2\sqrt{Q_0}}\pin{k}\int dx\partial_x\left(\omega_k \left|g_k(x)\right|^2+\frac{\left|g\p_k(x)\right|^2}{\omega_k}\right)=0.\nonumber\\
\eea
Note that although this is the integral of a total derivative, the differentiated function does not vanish at infinity.  The integral vanishes because the differentiated function is even in $x$.  This is true at each $k$ if the potential is symmetric under an inversion that exchanges the two minima responsible for the kink, as it is in the Sine-Gordon model and the $\phi^4$ model.   More generally, in the large $x$ region which fixes this integral by the fundamental theorem of calculus, the functions $g_k(x)$ are plane waves and their norm is constant and independent of the potential.  In the case of a reflectionless potential, the norm is equal in both asymptotic regimes and so this integral vanishes for each value of $k$.  More generally, the integral vanishes when summed over $k$ and $-k$ as the summed norms squared are equal in the two asymptotic regions.

Using this identity, one evaluates the contribution of $\vac_1^{21}$ to be
\beq
H_3\vac_1^{21}\supset\frac{1}{4\sqrt{Q_0}}\pin{k}\pin{k\p}V_{Bk-k\p}\Delta_{k\p B}\phi_0^3\B{}\vac_0.
\eeq
 Similarly
 \bea
H_3\vac_1^{12}&\supset&\frac{1}{4\sqrt{Q_0}}\pin{k}\pin{k\p}\left(\frac{\ok{}}{\okp{}}-1\right)V_{BB-k\p}\Delta_{kk\p}\phi_0^3\B{}\vac_0\\
\pin{k\p}\okp{}\Bp{}B_{k\p}\vac_2^{31}&=&\frac{1}{12Q_0}\pin{k}\pin{k\p}\left(3\okp{}-\ok{}\right)\ok{}\Delta_{-k\p B}\Delta_{k k\p}\phi_0^3\B{}\vac_0.\nonumber
\eea
 The final contribution is
 \beq
 H_4\vac_0=\frac{1}{6}\pin{k}V_{BBBk}\phi_0^3\B{}\vac_0.
 \eeq
 The $V$'s may all be traded for $\Delta$'s using (\ref{vid}) and, as may be derived similarly to (\ref{vbbbb}),
 \bea
 V_{BBBk}&=&\frac{1}{\sqrt{Q_0}}\int dx \partial_x\left(\V3\right)g_B(x)^2g_k(x)\label{vbbbk}\\
 &=&-\frac{1}{\sqrt{Q_0}}\int dx \V3\nonumber\\
 &&\times\left[2g_B(x)g_k(x)\int dy\left( g_B(x)g_B(y)+\pin{k\p}g_{k\p}(x)g_{-k\p}(y)\right)g_B\p(y)\right.\nonumber\\
 &&\left.+g^2_B(x)\int dy\left( g_B(x)g_B(y)+\pin{k\p}g_{k\p}(x)g_{-k\p}(y)\right)g_k\p(y)\right]\nonumber\\
 &=&-\frac{1}{\sqrt{Q_0}}\pin{k\p}\left(2V_{Bkk\p}\Delta_{-k\p B}+V_{BBk\p}\Delta_{-k\p k}\right)=\frac{1}{Q_0}\pin{k\p}\left(2\ok{}^2-3\okp{}^2\right)\Delta_{kk\p}\Delta_{-k\p B}.
\nonumber
 \eea
 Combining these contributions
 \bea
 \Gamma_2^{31}(k)&=&\pin{k\p}\Delta_{kk\p}\Delta_{-k\p B}\left(\frac{\okp{}^2-\ok{}^2}{4}-\frac{\okp{}^2}{4}\left(\frac{\ok{}}{\okp{}}-1\right)+\frac{3\okp{}\ok{}-\ok{}^2}{12}+\frac{2\ok{}^2-3\okp{}^2}{6}\right)\nonumber\\
 &=&0.
 \eea
 
 \subsubsection{$m=2,\ n=4$}
 The contributions are
 \bea
 H_3\vac_1^{21}&\supset&\frac{1}{12\sqrt{Q_0}}\pink{4}\ok{1}\Delta_{k_1B}V_{k_2k_3k_4}\phi_0^2\B1\B2\B3\B4\vac_0\\
H_3\vac_1^{12}&\supset&\frac{1}{4\sqrt{Q_0}}\pink{4}\left(\ok{1}-\ok{2}\right)\Delta_{k_1k_2}V_{Bk_3k_4}\phi_0^2\B1\B2\B3\B4\vac_0\nonumber\\
H_3\vac_1^{03}&\supset&-\frac{1}{12}\pink{4}V_{BBk_1}Y_{k_2k_3k_4}\phi_0^2\B1\B2\B3\B4\vac_0\nonumber\\
 \eea
 and
 \beq
  \pin{k\p}\okp{}\Bp{}B_{k\p}\vac_2^{24}=\pink{4}
  \left(\frac{Y_{Bk_1k_2}V_{Bk_3k_4}}{4}-\Omega_4\frac{\ok{1}\Delta_{k_1B}Y_{k_2k_3k_4}}{12\sqrt{Q_0}}  \right)
  \phi_0^2\B1\B2\B3\B4\vac_0.
  \eeq
 Therefore 
 \bea
 \Gamma_2^{24}(k_1\cdots k_4)&=&\frac{\sqrt{Q_0}\Delta_{k_1B}V_{k_2k_3k_4}}{12}\left[\ok{1}+\frac{\ok{1}^2}{\ok{2}+\ok{3}+\ok{4}}-\frac{\ok{1}\Omega_4}{\ok{2}+\ok{3}+\ok{4}}\right]\nonumber\\
 &&+\frac{\sqrt{Q_0}\Delta_{k_1k_2}V_{Bk_3k_4}}{4}\left[\left(\ok{1}-\ok{2}\right)+\frac{\ok{2}^2-\ok{1}^2}{\ok{1}+\ok{2}}\right]=0
 \eea
 as the terms in each square bracket vanish.
 
 \subsubsection{$m=2,\ n=2$}
 
 From here on there will be many more contributions to each $\Gamma_2^{mn}$, and so we will decompose them into pieces that are not expected to mix as was done in Subsec.~\ref{gamma0}.  First let us consider contributions that depend on $\I(x)$ and so on our renormalization scheme.  As we have seen that $V_{\I B}$ vanishes, there are three contributions
\bea
H_3\vac_1^{21}&\supset&\frac{1}{4\sqrt{Q_0}}\pink{2}V_{\I k_1}\ok{2}\Delta_{k_2 B}\phi_0^2\B1\B2\vac_0\\
H_3\vac_1^{01}&\supset&\frac{1}{4\sqrt{Q_0}}\pink{2}Y_{\I k_1}\ok{2}^2\Delta_{k_2 B}\phi_0^2\B1\B2\vac_0\nonumber\\
 \pin{k\p}\okp{}\Bp{}B_{k\p}\vac_2^{22}&\supset&-\frac{1}{4\sqrt{Q_0}}\pink{2}\left(\ok{1}+\ok{2}\right)Y_{\I k_1}\ok{2}\ok{2}^2\Delta_{k_2 B}\phi_0^2\B1\B2\vac_0\nonumber
\eea
 whose sum is easily seen to vanish.
 
Similarly to (\ref{vbbbk}) one may derive
\bea
V_{BBk_1k_2}&=&\frac{1}{Q_0}\left[-\left(\ok1^2+\ok2^2\right)\Delta_{k_1B}\Delta_{k_2B}\right.\\
&&\left.+\pin{k\p}\left[-\sqrt{Q_0}V_{k_1k_2k\p}\Delta_{-k\p B}+\left(\ok1^2+\ok2^2-2\okp{}^2\right)\Delta_{k_2k\p}\Delta_{-k\p k_1}\right]\right].\nonumber
\eea
 There are four terms that contain $V_{k_1k_2k_3}$
 \bea
 H_4\vac_0&\supset&-\frac{1}{4\sqrt{Q_0}}\pink{2}\pin{k\p}V_{k_1k_2k\p}\Delta_{-k\p B}\phi_0^2\B1\B2\vac_0\\
 H_3\vac_1^{03}&\supset&\frac{1}{8\sqrt{Q_0}}\pink{2}\pin{k\p}Y_{k_1k_2k\p}\okp{}\Delta_{-k\p B}\phi_0^2\B1\B2\vac_0\nonumber\\
  H_3\vac_1^{21}&\supset&\frac{1}{8\sqrt{Q_0}}\pink{2}\pin{k\p}V_{k_1k_2k\p}\Delta_{-k\p B}\phi_0^2\B1\B2\vac_0\nonumber\\
  \pin{k\p}\okp{}\Bp{}B_{k\p}\vac_2^{22}&\supset&\frac{1}{8\sqrt{Q_0}}\pink{2}\pin{k\p}\left(\ok{1}+\ok{2}\right)Y_{k_1k_2k\p}\Delta_{-k\p B}\phi_0^2\B1\B2\vac_0\nonumber
 \eea
which again sum to zero, as the second plus the fourth and also the third are equal to minus one half of the first.  The four terms with no $k\p$ integral are
\bea
H_4\vac_0&\supset&-\frac{1}{4Q_0}\pink{2}\left(\ok{1}^2+\ok{2}^2\right)\Delta_{k_1 B}\Delta_{k_2B}\phi_0^2\B1\B2\vac_0\\
H_3\vac_1^{01}&\supset&-\frac{1}{8Q_0}\pink{2}\left(\ok{1}^2+\ok{2}^2\right)\Delta_{k_1 B}\Delta_{k_2B}\phi_0^2\B1\B2\vac_0\nonumber\\
\frac{\pi_0^2}{2}\vac_2^{42}&=&-\frac{3}{4Q_0}\pink{2}\ok{1}\ok{2}\Delta_{k_1B}\Delta_{k_2B}\phi_0^2\B1\B2\vac_0\nonumber\\
 \pin{k\p}\okp{}\Bp{}B_{k\p}\vac_2^{22}&\supset&\frac{3}{8Q_0}\pink{2}\Omega_2^2\Delta_{k_1B}\Delta_{k_2B}\phi_0^2\B1\B2\vac_0\nonumber
 \eea
 which sum easily to zero as well.  Finally the three terms with $k\p$ but no $V_{k_1k_2k_3}$ are
 \bea
 H_4\vac_0&\supset&\frac{1}{4{Q_0}}\pink{2}\pin{k\p}\left(\ok{1}^2+\ok{2}^2-2\okp{}^2\right)\Delta_{k_1 k\p}\Delta_{-k\p k_2}\phi_0^2\B1\B2\vac_0\\
 H_3\vac_1^{12}&\supset&\frac{1}{2{Q_0}}\pink{2}\pin{k\p}\left(\frac{\ok{2}^2}{\okp{}}-\okp{}\right)\left(\ok{1}-\okp{}\right)\Delta_{k_1 k\p}\Delta_{-k\p k_2}\phi_0^2\B1\B2\vac_0\nonumber\\
 \pin{k\p}\okp{}\Bp{}B_{k\p}\vac_2^{22}&\supset&-\frac{1}{4Q_0}\pink{2}\pin{k\p}\Omega_2\left(\frac{\ok{1}\ok{2}}{\okp{}}-\okp{}\right)\Delta_{k_1 k\p}\Delta_{-k\p k_2}\phi_0^2\B1\B2\vac_0.\nonumber
 \eea
 Notice that in the second term, all factors are symmetric with respect to $k_1\leftrightarrow k_2$ except for the factors of $\omega$.  Therefore these may be symmetrized to
 \beq
 \frac{1}{2}\left(\frac{\ok{1}\ok{2}\Omega_2}{\okp{}}-\ok{1}^2-\ok{2}^2-\okp{}\Omega_2\right)+\okp{}^2
 \eeq
 which exactly cancels the corresponding contributions from the first and third terms.  We thus conclude that $\Gamma_2^{22}=0$.
 
 \subsubsection{$m=2,\ n=0$}
 We will see that this is the most interesting case so far, because it is the first that strongly depends on the form of $\I(x)$.  To see this, let us try to proceed as above.  The terms that depend on $\I(x)$ are
 \bea
 H_4\vac_0&\supset&\frac{1}{4}V_{\I BB}\phi_0^2\vac_0\\
 H_3\vac_1^{21}&\supset&\frac{1}{8\sqrt{Q_0}}\pin{k\p}V_{\I k\p}\Delta_{-k\p k_2}\phi_0^2\vac_0\nonumber\\
 H_3\vac_1^{01}&\supset&\frac{1}{8\sqrt{Q_0}}\pin{k\p}V_{\I k\p}\Delta_{-k\p k_2}\phi_0^2\vac_0.\nonumber
 \eea
 As above, we may eliminate $V^{(4)}$ using integration by parts and then inserting the completeness relation (\ref{comp})
 \bea
 V_{\I BB}&=&\int dx \V4 \I(x) g_B^2(x)=\frac{1}{\sqrt{Q_0}}\int dx \left(\partial_x\V3\right)\I(x)g_B(x)\\
 &=&-\frac{1}{\sqrt{Q_0}}\int dx \V3\left(\I\p(x)g_B(x)+\I(x)g\p_B(x)\right)\nonumber\\
 &=&-\frac{1}{\sqrt{Q_0}}\left(V_{\I\p B}+\pin{k\p}V_{\I k\p}\Delta_{-k\p B}\right).\nonumber
  \eea
 The second term cancels the contributions from $H_3\vac_1^{21}$ and $H_3\vac_1^{01}$, leaving
 \beq
 \Gamma_{2,\I}^{20}=-\frac{\sqrt{Q_0}}{4}V_{\I\p B}. \label{g20}
 \eeq
Let us rewrite $V_{\I\p B}$ in terms of quantities that we expect to find in other contributions.

Writing the identity (\ref{di}) as
\beq
\partial_x \I(x)=\pin{k}\frac{1}{2\ok{}}\left(g_k(x)g\p_{-k}(x)+g\p_k(x)g_{-k}(x)\right)
\eeq
one finds
\bea
V_{\I\p B}&=&\pin{k}\frac{1}{2\ok{}}\int dx \V3 g_B(x)\left(g_k(x)g\p_{-k}(x)+g\p_k(x)g_{-k}(x)\right)\\
&=&\pin{k}\frac{1}{\ok{}}\left(V_{BBk}\Delta_{Bk}+\pin{k\p}V_{Bkk\p}\Delta_{-k\p k}\right)\nonumber\\
&=&-\frac{1}{\sqrt{Q_0}}\pinkp{2}\frac{\okp{1}^2-\okp{2}^2}{\okp{1}} \Delta_{k\p_1k\p_2}\Delta_{-k\p_1-k\p_2}-\frac{1}{\sqrt{Q_0}}\pin{k\p}\okp{}\Delta_{Bk\p}\Delta_{-k\p B}.\nonumber
\eea
 Inserting this into (\ref{g20}) and symmetrizing dummy indices one finds
\beq
 \Gamma_{2,\I}^{20}=\frac{1}{8}\pinkp{2}\left(-\okp{1}-\okp{2}+\frac{\okp{1}^2}{\okp{2}}+\frac{\okp{2}^2}{\okp{1}}\right)
 \Delta_{k\p_1k\p_2}\Delta_{-k\p_1-k\p_2}+\frac{1}{4}\pin{k\p}\okp{}\Delta_{Bk\p}\Delta_{-k\p B}. \label{g20i}
\eeq
 Schrodinger's equation will only be satisfied if these terms are canceled by contributions with no $\I(x)$.
 
 There is only one contribution with no $\I(x)$ that has two contracted momenta $k\p$
 \beq
 H_3\vac_1^{12}\supset\frac{1}{8Q_0}\pinkp{2} \left(\okp{1}+\okp{2}-\frac{\okp{1}^2}{\okp{2}}-\frac{\okp{2}^2}{\okp{1}}\right) \Delta_{k\p_1k\p_2}\Delta_{-k\p_1-k\p_2}\phi_0^2\vac_0
 \eeq
 which indeed cancels the first term in (\ref{g20i}).  There are two contributions with no $\I(x)$ and a single $k\p$
 \bea
 H_3\vac_1^{01}&\supset&\frac{1}{8Q_0}\pin{k\p}\okp{}\Delta_{Bk\p}\Delta_{-k\p B}\phi_0^2\vac_0\\
 \frac{\pi_0^2}{2}\vac_2^{40}&=&-\frac{3}{8Q_0}\pin{k\p}\okp{}\Delta_{Bk\p}\Delta_{-k\p B}\phi_0^2\vac_0\nonumber
 \eea
 which are equal to $1/2$ and $-3/2$ of the second term in (\ref{g20i}), and so altogether they cancel, leaving $\Gamma_2^{20}=0$. 
 
 \subsubsection{$m=1,\ n=5$}
 There are only three contributions to this term
 \bea
 H_3\vac_1^{03}&\supset&\frac{1}{12\sqrt{Q_0}}\pink{5}\left(\ok{4}^2-\ok{5}^2\right)Y_{k_1k_2k_3}\Delta_{k_4k_5}\phi_0\B1\cdots \B5\vac_0\nonumber\\
 H_3\vac_1^{12}&\supset&\frac{1}{12\sqrt{Q_0}}\pink{5}\left(\ok{4}-\ok{5}\right)V_{k_1k_2k_3}\Delta_{k_4k_5}\phi_0\B1\cdots \B5\vac_0\nonumber\\
 \pin{k\p}\okp{}\Bp{}B_{k\p}\vac_2^{15}&=&-\frac{1}{12\sqrt{Q_0}}\pink{5}\left(\ok{4}-\ok{5}+\frac{\ok{4}^2-\ok{5}^2}{\Omega_3}\right)\nonumber\\&&\times V_{k_1k_2k_3}\Delta_{k_4k_5}\phi_0\B1\cdots \B5\vac_0
\eea
whose sum is readily seen to vanish. 
 
 \subsubsection{$m=1,\ n=3$}
 Again let us divide the 11 contributions to $\Gamma_2^{13}$ into 3 subsets that are expected to cancel separately.  First, terms involving $\I(x)$ are
 \bea
 H_3\vac_1^{12}&\supset&\frac{1}{4\sqrt{Q_0}}\pink{3}\left(\ok{2}-\ok{3}\right)V_{\I k_1}\Delta_{k_2k_3}\phi_0\B1\B2\B3\vac_0\\
  H_3\vac_1^{01}&\supset&\frac{1}{4\sqrt{Q_0}}\pink{3}\left(\ok{2}^2-\ok{3}^2\right)Y_{\I k_1}\Delta_{k_2k_3}\phi_0\B1\B2\B3\vac_0\nonumber\\
 \pin{k\p}\okp{}\Bp{}B_{k\p}\vac_2^{13}&\supset&-\frac{1}{4\sqrt{Q_0}}\pink{3}\left(\ok{2}-\ok{3}\right)\Omega_3Y_{\I k_1}\Delta_{k_2k_3}\phi_0\B1\B2\B3\vac_0\nonumber
 \eea
 which sum to zero.  
 
We will now need 
\bea
V_{Bk_1k_2k_3}&=&\frac{1}{Q_0}\left[\left(\ok2^2-\ok3^2\right)\Delta_{Bk_1}\Delta_{k_2k_3}+\left(\ok1^2-\ok3^2\right)\Delta_{Bk_2}\Delta_{k_1k_3}+\left(\ok1^2-\ok2^2\right)\Delta_{Bk_3}\Delta_{k_1k_2}\right]\nonumber\\
&&-\frac{1}{\sqrt{Q_0}}\pin{k\p}\left[V_{k_2k_3k\p}\Delta_{-k\p k_1}+V_{k_1k_3k\p}\Delta_{-k\p k_2}+V_{k_1k_2k\p}\Delta_{-k\p k_3}\right].
\eea
The terms which have a contracted index $k\p$ are
\bea
 H_4\vac_0&\supset&-\frac{1}{2\sqrt{Q_0}}\pink{3}\pin{k\p}V_{k_1k_2k\p}\Delta_{-k\p k_3}\phi_0\B1\B2\B3\vac_0\\
H_3\vac_1^{03}&\supset&\frac{1}{4\sqrt{Q_0}}\pink{3}\pin{k\p}\frac{\okp{}^2-\ok{3}^2}{\okp{}}Y_{k_1k_2k\p}\Delta_{-k\p k_3}\phi_0\B1\B2\B3\vac_0\nonumber\\
H_3\vac_1^{12}&\supset&\frac{1}{4\sqrt{Q_0}}\pink{3}\pin{k\p}\frac{\okp{}-\ok{3}}{\okp{}}V_{k_1k_2k\p}\Delta_{-k\p k_3}\phi_0\B1\B2\B3\vac_0\nonumber\\
\pin{k\p}\okp{}\Bp{}B_{k\p}\vac_2^{13}&\supset&\frac{1}{4\sqrt{Q_0}}\pink{3}\pin{k\p}\Omega_3\frac{\okp{}+\ok{3}}{\okp{}}Y_{k_1k_2k\p}\Delta_{-k\p k_3}\phi_0\B1\B2\B3\vac_0\nonumber
\eea
 whose sum also vanishes.  Finally the terms with neither $\I(x)$ nor $k\p$ are
 \bea
 H_4\vac_0&\supset&\frac{1}{2Q_0}\pink{3}\left(\ok{2}^2-\ok{3}^2\right)\Delta_{Bk_1}\Delta_{k_2 k_3}\phi_0\B1\B2\B3\vac_0\\
H_3\vac_1^{01}&\supset&\frac{1}{4Q_0}\pink{3}\left(\ok{2}^2-\ok{3}^2\right)\Delta_{Bk_1}\Delta_{k_2 k_3}\phi_0\B1\B2\B3\vac_0\nonumber\\
\frac{\pi_0^2}{2}\vac_2^{33}&\supset&\frac{3}{4Q_0}\pink{3}\ok{1}\left(\ok{2}-\ok{3}\right)\Delta_{Bk_1}\Delta_{k_2 k_3}\phi_0\B1\B2\B3\vac_0\nonumber\\
\pin{k\p}\okp{}\Bp{}B_{k\p}\vac_2^{13}&\supset&-\frac{3}{4Q_0}\pink{3}\left[\ok{1}\left(\ok{2}-\ok{2}\right)+\ok{2}^2-\ok{3}^2\right]\nonumber\\
&&\times\Delta_{Bk_1}\Delta_{k_2 k_3}\phi_0\B1\B2\B3\vac_0\nonumber
\eea
 which again trivially cancel, leaving  $\Gamma_2^{13}=0$. 
 
\subsubsection{$m=1,\ n=1$}
Finally we turn our attention to  $\Gamma_2^{11}.$  Like $\Gamma_2^{20}$ we will see that it only vanishes if $\I(x)$ satisfies (\ref{di}).  The terms involving $\I(x)$ are
\bea
 H_4\vac_0&\supset&\frac{1}{2}\pin{k}V_{\I Bk}\phi_0\B{}\vac_0\label{g11i}\\
 H_3\vac_1^{12}&\supset&\frac{1}{4\sqrt{Q_0}}\pin{k}\pin{k\p}\left(\okp{}-\ok{}\right)Y_{\I k\p}\Delta_{-k\p k}\phi_0\B{}\vac_0\nonumber\\
  H_3\vac_1^{01}&\supset&\frac{1}{4\sqrt{Q_0}}\pin{k}\pin{k\p}\left(\frac{\okp{}^2-\ok{}^2}{\okp{}}\right)Y_{\I k\p}\Delta_{-k\p k}\phi_0\B{}\vac_0\nonumber\\
\pin{k\p}\okp{}\Bp{}B_{k\p}\vac_2^{11}&\supset&\frac{1}{4\sqrt{Q_0}}\pin{k}\pin{k\p}\ok{}\left(\frac{\okp{}+\ok{}}{\okp{}}\right)Y_{\I k\p}\Delta_{-k\p k}\phi_0\B{}\vac_0.\nonumber
\eea
Again, as in the case of $\Gamma_2^{20}$ integration by parts allows us to remove the fourth derivative
\beq
V_{\I Bk}=-\frac{1}{\sqrt{Q_0}}\int dx \V3 \left(\I\p(x)g_k(x)+\I(x)g\p_k(x)\right)
\eeq
and the two terms can be simplified using completeness (\ref{comp}) and the formula (\ref{di}) for $\I\p(x)$
\bea
-\frac{1}{\sqrt{Q_0}}\int dx \V3 \I\p(x)g_k(x)&=&\frac{1}{Q_0}\pin{k\p}\left(\frac{\okp{}^2-\ok{}^2}{\okp{}}\right)\Delta_{kk\p}\Delta_{-k\p B}\label{vibk}\\
&&-\frac{1}{2\sqrt{Q_0}}\pinkp{2}\left(\frac{\okp{1}-\okp{2}}{\okp{1}\okp{2}}\right)V_{kk\p_1k\p_2}\Delta_{-k\p_1 -k\p_2}
\nonumber\\
-\frac{1}{\sqrt{Q_0}}\int dx \V3 \I(x)g\p_k(x)&=&-\frac{1}{\sqrt{Q_0}}V_{\I k\p}\Delta_{-k\p k}.\nonumber
\eea
 The second equation in (\ref{vibk}) substituted into the first term in (\ref{g11i}) cancels the second, third and fourth terms.  This leaves only the first equation in (\ref{vibk}), which when substituted into (\ref{g11i}) yields
 \beq
 \Gamma_{2,\I}^{11}=\frac{1}{2}\pin{k\p}\left(\frac{\okp{}^2-\ok{}^2}{\okp{}}\right)\Delta_{kk\p}\Delta_{-k\p B}-\frac{\sqrt{Q_0}}{4}\pinkp{2}\left(\frac{\okp{1}-\okp{2}}{\okp{1}\okp{2}}\right)V_{kk\p_1k\p_2}\Delta_{-k\p_1 -k\p_2}.
 \eeq

Summing the three contributions with integrals over $k\p_1$ and $k\p_2$
\bea
 H_3\vac_1^{03}&\supset&\frac{1}{8\sqrt{Q_0}}\pin{k}\pinkp{2}\left(\frac{\okp{1}^2-\okp{2}^2}{\okp{1}\okp{2}}\right)Y_{kk\p_1k\p_2}\Delta_{-k\p_1 -k\p_2}\phi_0\B{}\vac_0\nonumber\\
 H_3\vac_1^{12}&\supset&\frac{1}{8\sqrt{Q_0}}\pin{k}\pinkp{2}\left(\frac{\okp{1}-\okp{2}}{\okp{1}\okp{2}}\right)V_{kk\p_1k\p_2}\Delta_{-k\p_1 -k\p_2}\phi_0\B{}\vac_0\nonumber\\
\pin{k\p}\okp{}\Bp{}B_{k\p}\vac_2^{11}&\supset&\frac{1}{8\sqrt{Q_0}}\pin{k}\pinkp{2}\ok{1}\left(\frac{\okp{1}-\okp{2}}{\okp{1}\okp{2}}\right)Y_{kk\p_1k\p_2}\Delta_{-k\p_1 -k\p_2}\phi_0\B{}\vac_0\nonumber
\eea
 one obtains
\beq
 \Gamma_{2,2k\p}^{11}=\frac{\sqrt{Q_0}}{4}\pinkp{2}\left(\frac{\okp{1}-\okp{2}}{\okp{1}\okp{2}}\right)V_{kk\p_1k\p_2}\Delta_{-k\p_1 -k\p_2}
 \eeq
 which cancels the second term in $ \Gamma_{2,\I}^{11}$.
 
 Finally the terms with no $\I(x)$ and one $k\p$ are
 \bea
 H_3\vac_1^{01}&\supset&\frac{1}{4Q_0}\pin{k}\pin{k\p}\left(\okp{}-\frac{\ok{}^2}{\okp{}}\right)\Delta_{kk\p}\Delta_{-k\p B}\phi_0\B{}\vac_0\\
 \frac{\pi_0^2}{2}\vac_2^{31}&\supset&\frac{1}{4Q_0}\pin{k}\pin{k\p}\left(\ok{}-3\okp{}\right)\Delta_{kk\p}\Delta_{-k\p B}\phi_0\B{}\vac_0\nonumber\\
 \pin{k\p}\okp{}\Bp{}B_{k\p}\vac_2^{11}&\supset&\frac{1}{4Q_0}\pin{k}\pin{k\p}\left(-\ok{}+3\frac{\ok{}^2}{\okp{}}\right)\Delta_{kk\p}\Delta_{-k\p B}\phi_0\B{}\vac_0\nonumber
 \eea
 which sum to
 \beq
  \Gamma_{2,1k\p}^{11}=\frac{1}{2}\pin{k\p}\left(\frac{\ok{}^2-\okp{}^2}{\okp{}}\right)\Delta_{kk\p}\Delta_{-k\p B}
 \eeq
 canceling the first term  in $ \Gamma_{2,\I}^{11}$.  Summarizing, we have verified that
 \beq
 \Gamma_{2}^{11}=\Gamma_{2,\I}^{11}+\Gamma_{2,1k\p}^{11}+\Gamma_{2,2k\p}^{11}=0
 \eeq
 and so the state $\vac_2$ that we have found indeed solves Schrodinger's equation at two loops.
 
 \section{The Meson Mass in the Sine-Gordon Model} \label{mapp}
 In this Appendix we briefly review the Schrodinger picture derivation of the two-loop meson mass in the normal-ordered Sine-Gordon model.  First one expands the scalar field in terms of Heisenberg operators
\beq
\phi(x)=\pin{p}\left(A^\dag_p+\frac{A_{-p}}{2\omega_p}\right)e^{-ipx}.
\eeq 
The Sine-Gordon potential
\beq
V(x)=\frac{M^2}{g^2}\left(1-:{\rm{cos}}\left(g\phi(x)\right):_a\right)
\eeq
at fourth order is the interaction
\beq
H_4=-\frac{M^2g^2}{24}\int dx :\phi^4(x):_a
\eeq
while the free Hamiltonian is
\beq
H_2=\pin{p}\omega_p A^\dag_p A_p.
\eeq

 Let the meson state $|p\rangle$ be an eigenstate of the full Hamiltonian.  Expand it in powers of $g$
 \beq
 |p\rangle=\sum_{n=0}^{\infty}|p\rangle_n
 \eeq
 where
 \beq
 |p\rangle_0=A^\dag_p|\Omega\rangle.
 \eeq
 
 The tree level Schrodinger Equation
 \beq
 H_2|p\rangle_0=E_0|p\rangle_0
 \eeq
 is solved by $E_0=\omega_p$.  At one loop
 \beq
 0=(H_4-E_1)|p\rangle_0+(H_2-E_0)|p\rangle_1
 \eeq
 together with the convention\footnote{For simplicity we have fixed $p$ and applied a $p$-dependent normalization condition on $|\Omega\rangle$.}
 \beq
{}_0 \langle p|p\rangle_i=\delta_{0i} \label{ort}
\eeq
are solved by $E_1=0$ and
\bea
|p\rangle_1&=&\frac{M^2g^2}{24}\int dx\pinq{4}\frac{e^{-ix\sum_j^4 q_j}}{\sum_j^4 \omega_{q_j}}A^\dag_{q_1}\cdots A^\dag_{q_4}A^\dag_p|\Omega\rangle\nonumber\\
&+&\frac{M^2g^2}{12}\int dx\pinq{3}\frac{e^{-ix\left(-p+\sum_j^3 q_j\right)}}{\omega_p\left(-\omega_p+\sum_j^3 \omega_{q_j}\right)}A^\dag_{q_1}\cdots A^\dag_{q_3}|\Omega\rangle .
\eea

 At two loops, the Schrodinger equation is
  \beq
 0=(H_6-E_2)|p\rangle_0+(H_4-E_1)|p\rangle_1+(H_2-E_0)|p\rangle_2.
 \eeq
Let us left multiply ${}_0\langle p|$ and use the orthogonality condition (\ref{ort}).  As $H_6$ is normal ordered, its matrix element vanishes and one finds
\beq
E_2={}_0\langle p|H_4|p\rangle_1=A+B
\eeq
 where
 \bea
 A&=&-\frac{M^4g^4}{48\omega_p}\pinq{2} \frac{\omega_{q_1}+\omega_{q_2}+\omega_{p-q_1-q_2}}{\omega_{q_1}\omega_{q_2}\omega_{p-q_1-q_2}\left[\left(\omega_{q_1}+\omega_{q_2}+\omega_{p-q_1-q_2}\right)^2-\omega_p^2\right]}\\
 B&=&-\frac{M^4g^4}{384}\pinq{3}\frac{1}{\omega_{q_1}\omega_{q_2}\omega_{q_3}\omega_{q_1+q_2+q_3}\left(\omega_{q_1+q_2+q_3}+\sum_i^3\omega_{q_i}\right)}.\nonumber
 \eea
 
 The infrared divergent term $B$ is equal to the two-loop energy of the vacuum state $|\Omega\rangle$ and so it does not contribute to the meson mass.  Therefore the two-loop meson mass correction $M_2$ is equal to $A$ evaluated at $p=0$
 \beq
 M_2=-\frac{M^3g^4}{48}\pinq{2} \frac{\omega_{q_1}+\omega_{q_2}+\omega_{q_1+q_2}}{\omega_{q_1}\omega_{q_2}\omega_{q_1+q_2}\left[\left(\omega_{q_1}+\omega_{q_2}+\omega_{q_1+q_2}\right)^2-M^2\right]}=-\frac{Mg^4}{768}
 \eeq
 in agreement with the pole mass in Ref.~\cite{vega}.

\section* {Acknowledgement}

\noindent
JE is supported by the CAS Key Research Program of Frontier Sciences grant QYZDY-SSW-SLH006 and the NSFC MianShang grants 11875296 and 11675223.   JE also thanks the Recruitment Program of High-end Foreign Experts for support.


\begin{thebibliography}{99}

\bibitem{dhn1}
R.~F.~Dashen, B.~Hasslacher and A.~Neveu,
``Nonperturbative Methods and Extended Hadron Models in Field Theory 1. Semiclassical Functional Methods,''
Phys. Rev. D \textbf{10} (1974), 4114
doi:10.1103/PhysRevD.10.4114


\bibitem{rebhan}
  A.~Rebhan and P.~van Nieuwenhuizen,
  ``No saturation of the quantum Bogomolnyi bound by two-dimensional supersymmetric solitons,''
  Nucl.\ Phys.\ B {\bf 508} (1997) 449
  doi:10.1016/S0550-3213(97)00625-1, 10.1016/S0550-3213(97)80021-1
 [hep-th/9707163].

\bibitem{bad}
  A.~Alonso Izquierdo, W.~Garcia Fuertes, M.~A.~Gonzalez Leon, M.~de la Torre Mayado, J.~Mateos Guilarte and J.~M.~Munoz Castaneda,
  ``Lectures on the mass of topological solitons,''
  hep-th/0611180.

\bibitem{nastase}
H.~Nastase, M.~A.~Stephanov, P.~van Nieuwenhuizen and A.~Rebhan,
``Topological boundary conditions, the BPS bound, and elimination of ambiguities in the quantum mass of solitons,''
Nucl. Phys. B \textbf{542} (1999), 471-514
doi:10.1016/S0550-3213(98)00773-1
[arXiv:hep-th/9802074 [hep-th]].

\bibitem{lit}
A.~Litvintsev and P.~van Nieuwenhuizen,
``Once more on the BPS bound for the SUSY kink,''
[arXiv:hep-th/0010051 [hep-th]].

\bibitem{local}
A.~S.~Goldhaber, A.~Litvintsev and P.~van Nieuwenhuizen,
``Local Casimir energy for solitons,''
Phys. Rev. D \textbf{67} (2003), 105021
doi:10.1103/PhysRevD.67.105021
[arXiv:hep-th/0109110 [hep-th]].

\bibitem{hepp}
  K.~Hepp,
  ``The Classical Limit for Quantum Mechanical Correlation Functions,''
  Commun.\ Math.\ Phys.\  {\bf 35} (1974) 265.
  doi:10.1007/BF01646348

\bibitem{stuart}
D.~M.~A.~Stuart,
``Hamiltonian quantization of solitons in the $\phi^4_{1+1}$ quantum field theory. I. The semiclassical mass shift,''
[arXiv:1904.02588 [math-ph]].

\bibitem{mekink}
J.~Evslin,
``Manifestly Finite Derivation of the Quantum Kink Mass,''
JHEP \textbf{11} (2019), 161
doi:10.1007/JHEP11(2019)161
[arXiv:1908.06710 [hep-th]].

\bibitem{sg}
H.~Guo and J.~Evslin,
``Finite derivation of the one-loop sine-Gordon soliton mass,''
JHEP \textbf{02} (2020), 140
doi:10.1007/JHEP02(2020)140
[arXiv:1912.08507 [hep-th]].



\bibitem{me2stato}
J.~Evslin,
``Constructing Quantum Soliton States Despite Zero Modes,''
[arXiv:2006.02354 [hep-th]].





\bibitem{cahill76}
K.~E.~Cahill, A.~Comtet and R.~J.~Glauber,
``Mass Formulas for Static Solitons,''
Phys. Lett. B \textbf{64} (1976), 283-285
doi:10.1016/0370-2693(76)90202-1


\bibitem{memassa}
J.~Evslin,
``Well-defined quantum soliton masses without supersymmetry,''
Phys. Rev. D \textbf{101} (2020) no.6, 065005
doi:10.1103/PhysRevD.101.065005
[arXiv:2002.12523 [hep-th]].

\bibitem{wick}
J.~Evslin,
``Normal Ordering Normal Modes,''
[arXiv:2007.05741 [hep-th]].



\bibitem{vega}
H.~J.~de Vega,
``Two-Loop Quantum Corrections to the Soliton Mass in Two-Dimensional Scalar Field Theories,''
Nucl. Phys. B \textbf{115} (1976), 411-428
doi:10.1016/0550-3213(76)90497-1

\bibitem{verwaest}
J.~Verwaest,
``Higher Order Correction to the Sine-Gordon Soliton Mass,''
Nucl. Phys. B \textbf{123} (1977), 100-108
doi:10.1016/0550-3213(77)90343-1


\bibitem{gjs}
J.~L.~Gervais, A.~Jevicki and B.~Sakita,
``Perturbation Expansion Around Extended Particle States in Quantum Field Theory. 1.,''
Phys. Rev. D \textbf{12} (1975), 1038
doi:10.1103/PhysRevD.12.1038


\bibitem{dhn2}
  R.~F.~Dashen, B.~Hasslacher and A.~Neveu,
  ``Nonperturbative Methods and Extended Hadron Models in Field Theory 2. Two-Dimensional Models and Extended Hadrons,''
  Phys.\ Rev.\ D {\bf 10} (1974) 4130.
 doi:10.1103/PhysRevD.10.4130


\bibitem{zamkink}
A.~B.~Zamolodchikov,
``Mass scale in the sine-Gordon model and its reductions,''
Int. J. Mod. Phys. A \textbf{10} (1995), 1125-1150
doi:10.1142/S0217751X9500053X



\bibitem{aguirre}
A.~R.~Aguirre and G.~Flores-Hidalgo,
``A note on one-loop soliton quantum mass corrections,''
Mod. Phys. Lett. A \textbf{35} (2020) no.13, 2050102
doi:10.1142/S0217732320501023
[arXiv:1912.13051 [hep-th]].

\bibitem{takyi}
I.~Takyi, M.~K.~Matfunjwa and H.~Weigel,
``Quantum Corrections to Solitons in the $\Phi^8$ Model,''
[arXiv:2010.07182 [hep-th]].

\bibitem{kinkfen}
M.~Lizunova and J.~van Wezel,
``An introduction to kinks in $\varphi^4$-theory,''
[arXiv:2009.00355 [nlin.PS]].

\bibitem{adamscat}
C.~Adam, K.~Oles, T.~Romanczukiewicz and A.~Wereszczynski,
``Kink-antikink collisions in a weakly interacting $\phi^4$ model,''
[arXiv:1912.09371 [hep-th]].

\bibitem{chris}
C.~Halcrow,
``Quantum soliton scattering manifolds,''
JHEP \textbf{07} (2020), 182
doi:10.1007/JHEP07(2020)182
[arXiv:2004.14167 [hep-th]].

\bibitem{wobble}
A.~Alonso-Izquierdo, L.~M.~Nieto and J.~Queiroga-Nunes,
``Scattering between wobbling kinks,''
[arXiv:2007.15517 [hep-th]].

\bibitem{melac}
I.~V.~Melnikov, C.~Papageorgakis and A.~B.~Royston,
``Accelerating Solitons,''
[arXiv:2007.11028 [hep-th]].

\bibitem{melac2}
I.~Melnikov, C.~Papageorgakis and A.~B.~Royston,
``The Forced Soliton Equation and Semiclassical Soliton Form Factors,''
[arXiv:2010.10381 [hep-th]].

\bibitem{royff}
A.~Roy, D.~Schuricht, J.~Hauschild, F.~Pollmann and H.~Saleur,
``The quantum sine-Gordon model with quantum circuits,''
[arXiv:2007.06874 [quant-ph]].

\bibitem{kimform}
J.~Y.~Kim and B.~D.~Sun,
``Gravitational form factors of a baryon with spin-3/2,''
[arXiv:2011.00292 [hep-ph]].

\end{thebibliography}

\begin{thebibliography}{99}

\bibitem{hepp}
  K.~Hepp,
  ``The Classical Limit for Quantum Mechanical Correlation Functions,''
  Commun.\ Math.\ Phys.\  {\bf 35} (1974) 265.
  doi:10.1007/BF01646348

\bibitem{sato}
J.~Sato and T.~Yumibayashi,
``Quantum-classical correspondence via coherent state in integrable field theory,''
[arXiv:1811.03186 [quant-ph]].

\bibitem{taylor78}
  J.~G.~Taylor,
  ``Solitons as Infinite Constituent Bound States,''
  Annals Phys.\  {\bf 115} (1978) 153.
  doi:10.1016/0003-4916(78)90179-3

\bibitem{delfino}
  G.~Delfino, W.~Selke and A.~Squarcini,
  ``Vortex mass in the three-dimensional $O(2)$ scalar theory,''
  Phys.\ Rev.\ Lett.\  {\bf 122} (2019) no.5,  050602
  doi:10.1103/PhysRevLett.122.050602
  [arXiv:1808.09276 [cond-mat.stat-mech]].

\bibitem{davies}
  D.~Davies,
  ``Quantum Solitons in any Dimension: Derrick's Theorem v. AQFT,''
  arXiv:1907.10616 [hep-th].


\bibitem{colemansg}
  S.~R.~Coleman,
  ``The Quantum Sine-Gordon Equation as the Massive Thirring Model,''
  Phys.\ Rev.\ D {\bf 11} (1975) 2088.
  doi:10.1103/PhysRevD.11.2088


\bibitem{mandelstamsol}
  S.~Mandelstam,
  ``Soliton Operators for the Quantized Sine-Gordon Equation,''
  Phys.\ Rev.\ D {\bf 11} (1975) 3026.
  doi:10.1103/PhysRevD.11.3026

\bibitem{sw2}
  N.~Seiberg and E.~Witten,
  ``Electric - magnetic duality, monopole condensation, and confinement in N=2 supersymmetric Yang-Mills theory,''
  Nucl.\ Phys.\ B {\bf 426} (1994) 19
   Erratum: [Nucl.\ Phys.\ B {\bf 430} (1994) 485]
  doi:10.1016/0550-3213(94)90124-4, 10.1016/0550-3213(94)00449-8
  [hep-th/9407087].

\bibitem{thooftconf}
  G.~'t Hooft,
  ``Topology of the Gauge Condition and New Confinement Phases in Nonabelian Gauge Theories,''
  Nucl.\ Phys.\ B {\bf 190} (1981) 455.
  doi:10.1016/0550-3213(81)90442-9

\bibitem{mandelconf}
  S.~Mandelstam,
  ``Vortices and Quark Confinement in Nonabelian Gauge Theories,''
  Phys.\ Rept.\  {\bf 23} (1976) 245.
  doi:10.1016/0370-1573(76)90043-0

\bibitem{rebsol}
A.~Aguirre and G.~Flores-Hidalgo,
``A note on one-loop soliton quantum mass corrections,''
Mod. Phys. Lett. A \textbf{33} (2020), 2050102
doi:10.1142/S0217732320501023
[arXiv:1912.13051 [hep-th]].

\bibitem{dhn2}
  R.~F.~Dashen, B.~Hasslacher and A.~Neveu,
  ``Nonperturbative Methods and Extended Hadron Models in Field Theory 2. Two-Dimensional Models and Extended Hadrons,''
  Phys.\ Rev.\ D {\bf 10} (1974) 4130.
 doi:10.1103/PhysRevD.10.4130

\bibitem{rajaraman}
  R.~Rajaraman,
  ``Some Nonperturbative Semiclassical Methods in Quantum Field Theory: A Pedagogical Review,''
  Phys.\ Rept.\  {\bf 21} (1975) 227.
  doi:10.1016/0370-1573(75)90016-2

\bibitem{physrept04}
  A.~S.~Goldhaber, A.~Rebhan, P.~van Nieuwenhuizen and R.~Wimmer,
  ``Quantum corrections to mass and central charge of supersymmetric solitons,''
  Phys.\ Rept.\  {\bf 398} (2004) 179
  doi:10.1016/j.physrep.2004.05.001
  [hep-th/0401152].

\bibitem{rebhan}
  A.~Rebhan and P.~van Nieuwenhuizen,
  ``No saturation of the quantum Bogomolnyi bound by two-dimensional supersymmetric solitons,''
  Nucl.\ Phys.\ B {\bf 508} (1997) 449
  doi:10.1016/S0550-3213(97)00625-1, 10.1016/S0550-3213(97)80021-1
 [hep-th/9707163].

\bibitem{dhn1}
R.~F.~Dashen, B.~Hasslacher and A.~Neveu,
``Nonperturbative Methods and Extended Hadron Models in Field Theory 1. Semiclassical Functional Methods,''
Phys. Rev. D \textbf{10} (1974), 4114
doi:10.1103/PhysRevD.10.4114

\bibitem{mestato}
J.~Evslin,
``The Ground State of the Sine-Gordon Soliton,''
[arXiv:2003.11384 [hep-th]].

\bibitem{friedrichscont}
K.O. Friedrichs,
``On the perturbation of continuous spectra,''
Commun.Pure Appl.Math. 1 (1948) 361-406
doi:10.1002/cpa.3160010404

\bibitem{callangross}
C.~G.~Callan, Jr. and D.~J.~Gross,
``Quantum Perturbation Theory of Solitons,''
Nucl. Phys. B \textbf{93} (1975), 29-55
doi:10.1016/0550-3213(75)90150-9

\bibitem{coleman2d}
S.~R.~Coleman,
``There are no Goldstone bosons in two-dimensions,''
Commun. Math. Phys. \textbf{31} (1973), 259-264
doi:10.1007/BF01646487

\bibitem{mekink}
J.~Evslin,
``Manifestly Finite Derivation of the Quantum Kink Mass,''
JHEP \textbf{11} (2019), 161
doi:10.1007/JHEP11(2019)161
[arXiv:1908.06710 [hep-th]].

\bibitem{memassa}
J.~Evslin,
``Well-defined quantum soliton masses without supersymmetry,''
Phys. Rev. D \textbf{101} (2020) no.6, 065005
doi:10.1103/PhysRevD.101.065005
[arXiv:2002.12523 [hep-th]].



\bibitem{sg}
H.~Guo and J.~Evslin,
``Finite derivation of the one-loop sine-Gordon soliton mass,''
JHEP \textbf{02} (2020), 140
doi:10.1007/JHEP02(2020)140
[arXiv:1912.08507 [hep-th]].


\bibitem{flugge}
S. Fl\"ugge,
``Practical Quantum Mechanics,"
Springer-Verlag Berlin Heidelberg (1999),
doi:10.1007/978-3-642-61995-3

\bibitem{vanhove1}
L.~Van Hove,
``Energy corrections and persistent perturbation effects in continuous spectra,''
Physica \textbf{21} (1955), 901-923
doi:10.1016/S0031-8914(55)92832-9  
  
\bibitem{vanhove2}
L.~Van Hove,
``Energy corrections and persistent perturbation effects in continuous spectra. II: The perturbed stationary states,''
Physica \textbf{22} (1956), 343-354
doi:10.1016/S0031-8914(56)80046-3

\bibitem{sakitacc}
J.~L.~Gervais and B.~Sakita,
``Extended Particles in Quantum Field Theories,''
Phys. Rev. D \textbf{11} (1975), 2943
doi:10.1103/PhysRevD.11.2943


\bibitem{stuart}
D.~Stuart, M.A.,
``Hamiltonian quantization of solitons in the $\phi^4_{1+1}$ quantum field theory. I. The semiclassical mass shift,''
[arXiv:1904.02588 [math-ph]].

\bibitem{firrotta1}
M.~Bianchi and M.~Firrotta,
``DDF operators, open string coherent states and their scattering amplitudes,''
Nucl. Phys. B \textbf{952} (2020), 114943
doi:10.1016/j.nuclphysb.2020.114943
[arXiv:1902.07016 [hep-th]].

\bibitem{firrotta2}
A.~Aldi and M.~Firrotta,
``String coherent vertex operators of Neveu-Schwarz and Ramond states,''
Nucl. Phys. B \textbf{955} (2020), 115050
doi:10.1016/j.nuclphysb.2020.115050
[arXiv:1912.06177 [hep-th]].


\bibitem{dhn2loops}
R.~F.~Dashen, B.~Hasslacher and A.~Neveu,
``The Particle Spectrum in Model Field Theories from Semiclassical Functional Integral Techniques,''
Phys. Rev. D \textbf{11} (1975), 3424
doi:10.1103/PhysRevD.11.3424

\bibitem{luther}
A.~Luther,
``Eigenvalue spectrum of interacting massive fermions in one-dimension,''
Phys. Rev. B \textbf{14} (1976), 2153-2159
doi:10.1103/PhysRevB.14.2153

\bibitem{vega}
H.~de Vega,
``Two-Loop Quantum Corrections to the Soliton Mass in Two-Dimensional Scalar Field Theories,''
Nucl. Phys. B \textbf{115} (1976), 411-428
doi:10.1016/0550-3213(76)90497-1

\bibitem{mussardo}
G.~Mussardo, V.~Riva and G.~Sotkov,
``Semiclassical scaling functions of sine-Gordon model,''
Nucl. Phys. B \textbf{699} (2004), 545-574
doi:10.1016/j.nuclphysb.2004.08.004
[arXiv:hep-th/0405139 [hep-th]].


\end{thebibliography}

\begin{thebibliography}{99}

\bibitem{dhn2}
  R.~F.~Dashen, B.~Hasslacher and A.~Neveu,
  ``Nonperturbative Methods and Extended Hadron Models in Field Theory 2. Two-Dimensional Models and Extended Hadrons,''
  Phys.\ Rev.\ D {\bf 10} (1974) 4130.
  doi:10.1103/PhysRevD.10.4130
  
\bibitem{dhnsg}
  R.~F.~Dashen, B.~Hasslacher and A.~Neveu,
  ``The Particle Spectrum in Model Field Theories from Semiclassical Functional Integral Techniques,''
  Phys.\ Rev.\ D {\bf 11} (1975) 3424.
  doi:10.1103/PhysRevD.11.3424


\bibitem{colemansg}
  S.~R.~Coleman,
  ``The Quantum Sine-Gordon Equation as the Massive Thirring Model,''
  Phys.\ Rev.\ D {\bf 11} (1975) 2088.
  doi:10.1103/PhysRevD.11.2088

\bibitem{luther}
  A.~Luther,
  ``Eigenvalue spectrum of interacting massive fermions in one-dimension,''
  Phys.\ Rev.\ B {\bf 14} (1976) 2153.
  doi:10.1103/PhysRevB.14.2153


\bibitem{re}
  A.Rebhan and P.Van Nieuwenhuizen,
  ``No saturation of the quantum Bogomolnyi bound by two-dimensional supersymmetric solitons,''
  [hep-th/9707163]

\bibitem{mekink}
 J.~Evslin,
  ``Manifestly Finite Derivation of the Quantum Kink Mass,''
  JHEP {\bf 1911} (2019) 161
  doi:10.1007/JHEP11(2019)161
  [arXiv:1908.06710 [hep-th]].

\bibitem{hui}
  H.~Liu, Y.~Zhou and J.~Evslin,
  ``Ground States of the $\phi^4$ Double-Well QFT,''
  arXiv:1909.04946 [hep-th].

\bibitem{taylor78}
  J.~G.~Taylor,
  ``Solitons as Infinite Constituent Bound States,''
  Annals Phys.\  {\bf 115} (1978) 153.
  doi:10.1016/0003-4916(78)90179-3


\bibitem{hepp}
  K.~Hepp,
  ``The Classical Limit for Quantum Mechanical Correlation Functions,''
  Commun.\ Math.\ Phys.\  {\bf 35} (1974) 265.
  doi:10.1007/BF01646348






\bibitem{flugge}
S. Fl\"ugge,
``Practical Quantum Mechanics,"
Springer-Verlag Berlin Heidelberg (1999),
doi:10.1007/978-3-642-61995-3

\bibitem{johnson73}
  J.~D.~Johnson, S.~Krinsky and B.~M.~McCoy,
  ``Vertical-Arrow Correlation Length in the Eight-Vertex Model and the Low-Lying Excitations of the X-Y-Z Hamiltonian,''
  Phys.\ Rev.\ A {\bf 8} (1973) 2526.

\bibitem{ft}
  L.~D.~Faddeev, L.~A.~Takhtajan and V.~E.~Zakharov,
  ``Complete description of solutions of the Sine-Gordon equation,''
  Dokl.\ Akad.\ Nauk Ser.\ Fiz.\  {\bf 219} (1974) 1334
   [Sov.\ Phys.\ Dokl.\  {\bf 19} (1975) 824].


\bibitem{mandelop}
  S.~Mandelstam,
  ``Soliton Operators for the Quantized Sine-Gordon Equation,''
  Phys.\ Rev.\ D {\bf 11} (1975) 3026.
  doi:10.1103/PhysRevD.11.3026

\bibitem{sw2}
  N.~Seiberg and E.~Witten,
  ``Electric - magnetic duality, monopole condensation, and confinement in N=2 supersymmetric Yang-Mills theory,''
  Nucl.\ Phys.\ B {\bf 426} (1994) 19
   Erratum: [Nucl.\ Phys.\ B {\bf 430} (1994) 485]
  doi:10.1016/0550-3213(94)90124-4, 10.1016/0550-3213(94)00449-8
  [hep-th/9407087].


\end{thebibliography}
\end{document}